\documentclass[manuscript=article, layout=twocolumn]{achemso}

\usepackage[version=3]{mhchem} 
\usepackage[]{braket} 
\usepackage[]{multirow} 
\usepackage{tikz}
\usepackage{lscape} 
\usepackage{longtable}
\usetikzlibrary{decorations.pathreplacing}
\usetikzlibrary{patterns}
\usetikzlibrary{shapes.geometric}
\usetikzlibrary{tikzmark}
\usepackage{pgfplots}
\usetikzlibrary{perspective,3d,patterns.meta}

\author{Laurenz Monzel}
 \affiliation{Fachrichtung Chemie, Universität des Saarlandes, Campus B2.2, D-66123 Saarbrücken, Germany}
\author{Stella Stopkowicz}
\affiliation{Fachrichtung Chemie, Universität des Saarlandes, Campus B2.2, D-66123 Saarbrücken, Germany}%
\email{stella.stopkowicz@uni-saarland.de}
\alsoaffiliation{Hylleraas Centre for Quantum Molecular Sciences, Department of Chemistry, University of Oslo, P.O. Box 1033 Blindern, N-0315 Oslo, Norway}

\title[Polaritonic Coupled Cluster Theory for Unpolarized Cavities Exploiting Point Group Symmetry ]{Polaritonic Coupled Cluster Theory for Unpolarized Cavities Exploiting Point Group Symmetry}
\abbreviations{IR,NMR,UV}

\SectionNumbersOn

\begin{document}

\begin{tocentry}

\end{tocentry}

\begin{abstract}
  We introduce a generalization of the quantum electrodynamic coupled cluster (QED-CC) wave function ansatz, to describe the strongly coupled light-matter system in an unpolarized optical Fabry-Pérot cavity.
  This is achieved by explicitly treating two cavity modes in our calculation with perpendicular polarizations and demonstrate that this ansatz preserves the symmetry of an unpolarized cavity. 
  Furthermore, exploiting point-group symmetry enables the assignment of polaritonic excited states as well as their targeted calculation. 
  Using our implementation, the aromatic species benzene, fluorobenzene and azulene are investigated. 
  We demonstrate that molecules in unpolarized cavities have a complicated excited-state landscapes with a plethora of avoided-crossings.
  We compare the results for a cavity with a single polarization to those of an unpolarized cavity described by two perpendicular polarization vectors using the excited states of the H$_2$ molecule as an example. 
\end{abstract}

\section{Introduction}
\label{sec:Introduction}
Manipulating molecular properties in a targeted manner is a central objective of chemistry. 
Historically, methods are sought that have the most specific influence while being easy to implement experimentally. 
One possibility that has received significant  attention in recent years, is the strong coupling to an electromagnetic field.\cite{ebbesen2016hybrid,thomas2016ground,wonmi,Walther2006,schwartz2011reversible}
By placing molecules in cavities, light-matter hybrids are formed, which can be manipulated by carefully choosing the frequency and coupling strength of the cavity.\cite{shalabney2015coherent,ruggenthaler2023understanding}
The strong coupling leads to the mixing of electronic and photonic degrees of freedom, forming so-called polaritons.\cite{herrera2016cavity,ebbesen2016hybrid,ribeiro2018polariton}
This enables the control of the matter states at room temperature with an relatively cheap experimental setup.\cite{wang2014phase}
Further, as the cavity mediated interaction between particles is not decaying with their distance, collective phenomena occur which are currently heavily investigated experimentally and theoretically.\cite{ebbesen2016hybrid,wonmi}
\\  
Regarding matter systems, a plethora of theoretical methods have been formulated which aim to solve for the many-particle wave function of the system. 
In order to also include the cavity, these methods have been generalized to also include effects emerging from quantum electrodynamics (QED).
Regarding optical cavities which strongly couple to electronic transitions, QED Hartree-Fock theory (QED-HF),\cite{haugland2020coupled,riso2024toward}  density functional theory (QEDFT)\cite{ruggenthaler2014quantum,flick2017atoms,flick_2018}, configuration interaction theory (QED-CI)\cite{PhysRevResearch.2.023262,haugland2021intermolecular}, perturbation theory (QED-PT)\cite{bauer2023perturbation,moutaoukal2025moller}, multi-reference methods (QED-CASSCF)\cite{vu2024cavity} and various flavors of coupled-cluster theory (QED-CC) \cite{mordovina2020polaritonic,haugland2020coupled,flick2021polaritonic,pathak2024quantum} have been formulated.
Compared to model Hamiltonians as the Dicke-, Jaynes- or Tavis-Cummings models,\cite{hertzog2019strong,ruggenthaler2018quantum} the aforementioned methods have the advantage that they allow for the relaxation of the electronic structure and treat electrons and photons at the same footing. 
In particular, it is mostly QED-CC theory and its variants which have been proven to provide highly accurate results. 
Recent milestones in polaritonic CC theory were the initial formulation,\cite{mordovina2020polaritonic,haugland2020coupled} treating circularly polarized cavities,\cite{riso2023strong} and the implementation of molecular gradients.\cite{lexander2024analytical}
\\
Yet, QED-CC theory has only been formulated for single polarizations, both for linearly and circularly polarized cavities.\cite{haugland2020coupled,riso2023strong,PhysRevResearch.2.023262}  
By going beyond the single-mode approximation and explicitly treating two perpendicularly polarized cavity modes, we enable the description of unpolarized cavities as often used experimentally, such as, for example, the Fabry-Pérot cavity.\cite{yokoyama}
\\
In this paper, we present a CC implementation for molecules in unpolarized cavities in the dipole-approximation\cite{di2019resolution,rokaj_2018} by explicitly treating two perpendicularly polarized modes. 
Further, we present how point-group symmetry can be applied for molecules in cavities and its exploitation in the speed-up of the respective calculations as well as in the interpretation of polaritonic states. 
An efficient implementation of point-group symmetry in CC calculations based on the direct-product decomposition\cite{stanton1991symmetry} is adapted here for the use in our QED-CC implementation. 
\\
First, in section (\ref{sec:theory}) the underlying theory of a strongly coupled light-matter system in an unpolarized Fabry-Pérot cavity is presented. 
We present further, how the direct-product decomposition is used for the evaluation of the non-linear coupled cluster (CC) equations.\cite{crawford2007introduction}
Section (\ref{sec:calculations}) presents exemplary calculations on molecules to study the effects of an unpolarized cavity for the ground and excited states. 
In section (\ref{sec:aromatics})-(\ref{subsec:azulene}), ground-state calculations on the aromatic species benzene, fluorobenzene as well as azulene are presented.
In Sections (\ref{subsec:excited_states})-(\ref{subsec:H2}) we present excited state calculations for unpolarized cavities with the help of EOM-CC and compare the coupling mechanism for a linearly and unpolarized cavity.

\section{Theory}
\label{sec:theory}
We will follow the index convention that $a,b,c,\cdots$ refer to virtual orbitals, $i,j,k,\cdots$ to occupied orbitals, and $p,q,r,\cdots$ are generic orbital indices. 
Photons are denoted with greek letters as $\alpha,\beta,\gamma,\cdots$ .
Further, we will express all operators and quantities in terms of atomic units 
\begin{equation}
  \hbar = e = m_\text{e} = a_0 = 1\;.
  \label{eq:atomicunits}
\end{equation}
Throughout, we stay in the polaritonic partitioning for the Born-Oppenheimer approximation and separate the nuclear motion from the electron-photon wave function.\cite{ruggenthaler2023understanding}

\subsection{The Dipole Hamiltonian}
\label{subsec:em_field}
In the dipole approximation it is assumed that the wavelength of the electromagnetic (EM) field exceeds the dimension of the molecule.
\cite{di2019resolution,rokaj_2018,flick_2018,ruggenthaler2023understanding}
The interacting system of electrons and the EM field is then derived from the Pauli-Fierz Hamiltonian in the length gauge which is in the coherent state basis given by\cite{ruggenthaler2023understanding}
\begin{equation}
	\begin{split}
      \hat{H} &= \hat{\tilde{H}}_\text{el} - \hat{\boldsymbol{D}}\cdot \hat{\tilde{\boldsymbol{d}}} + \hat{H}_\text{cav} .
	\end{split}
   \label{eq:ham}
\end{equation}
This Hamiltonian  describes the electrons interacting with homogeneously oscillating electric fields, where the light-matter interaction is captured by the product of the phase-shifted homogeneous displacement field $\hat{\boldsymbol{D}}$ and the dipole fluctuation operator $\hat{\tilde{\boldsymbol{d}}}$. 
The phase-shifted displacement field $\hat{\boldsymbol{D}}$ in the length gauge reads
\begin{equation}
\hat{\boldsymbol{D}} = \sum_\alpha^{N_\text{cav}} \sqrt{\frac{\lambda_\alpha^2 \omega_\alpha}{2}} \boldsymbol{\epsilon}_\alpha (\hat{\alpha}^\dagger + \hat{\alpha})
   \label{}
\end{equation}
where each mode $\alpha$ is characterized by its frequency $\omega_\alpha$, coupling strength $\lambda_\alpha$ and field polarizations $\boldsymbol{\epsilon}_\alpha$ (the field polarization is throughout assumed to be real-valued).  
The photon creation and annihilation operators are denoted by $\hat{\alpha}^\dagger$ and $\hat{\alpha}$.
In the coherent-state basis, $\hat{\tilde{\boldsymbol{d}}}$ is the dipole operator in normal order
\begin{equation}
   \hat{\tilde{\boldsymbol{d}}} = \hat{\boldsymbol{d}} - \braket{\boldsymbol{d}}
   \label{}
\end{equation}
with the dipole operator  $\hat{\boldsymbol{d}}$.\cite{haugland2020coupled,moutaoukal2025moller}
The operator $\hat{\tilde{\boldsymbol{d}}}$ describes the dipole fluctuation relative to the Fermi vacuum.\cite{shavitt2009many}
The coherent-state basis guarantees the translational invariance of the Hamiltonian. 
Typically, a Hartree-Fock (HF) reference determinant is chosen for the Fermi vacuum, while photons are described in a coherent-state ansatz.\cite{haugland2020coupled}
An alternative ansatz for the reference wave function could possibly be a strong-coupling QED Hartree-Fock (SC-QED-HF)\cite{riso2024toward} reference, but, so far this ansatz has only been exploited in a perturbation-theory framework.\cite{ moutaoukal2025moller}
The electronic Hamiltonian $\hat{H}_\text{el}$ includes the kinetic energy as well as the electrostatic electron-nuclei and electron-electron interactions 
\begin{equation}
  \hat{H}_\text{el} = \hat{T}_\text{e} + V_\text{eN} + W_\text{ee}  \;.
   \label{eq:Hel}
\end{equation}
The cavity Hamiltonian (including the matter system) is given by
\begin{equation}
\hat{H}_\text{cav} = \sum_\alpha^{N_\text{cav}} \Big( \omega_{\alpha} \hat{\alpha}^\dagger \hat{\alpha} + \frac{\lambda_\alpha^2}{2}  (\boldsymbol{\epsilon} \cdot \hat{\tilde{\boldsymbol{d}}} )^2 \Big) \;.
   \label{eq:Hph}
\end{equation}
The first term corresponds to the excitation of the displacement modes and the second term is the dipole self-energy contribution.
Note that the dipole self-energy term ensures that the Hamiltonian is bound from below. It hence enables the computation of polaritonic ground states.\cite{ruggenthaler2023understanding,rokaj_2018, flick_2018, Fabri2025impact}
Furthermore, we note that the dipole self-energy is part of the photon field, even though it is technically a purely electronic operator. 
In practical calculations, the summation over the cavity modes has to be truncated and only a few effective modes that lie close to electronic transitions are typically included.
Which modes are selected can heavily influence the light-matter coupling and symmetry of the system, which will be discussed in the following sections.
Note also, that when taking more modes into account, mass renormalization effects increase. \cite{svendsen2025}
Here, we only take two modes (with the same frequency) into account and assume that these effects are negligible. 
A procedure how to include mass renormalization effects can be found in Ref. \citenum{PhysRevResearch.7.013093}.

\subsection{Point-group Symmetry for the Bare Cavity Hamiltonian}
\label{sec:point-symmetry}
\begin{figure}[t]
    \centering
    \includegraphics[width=0.7\linewidth]{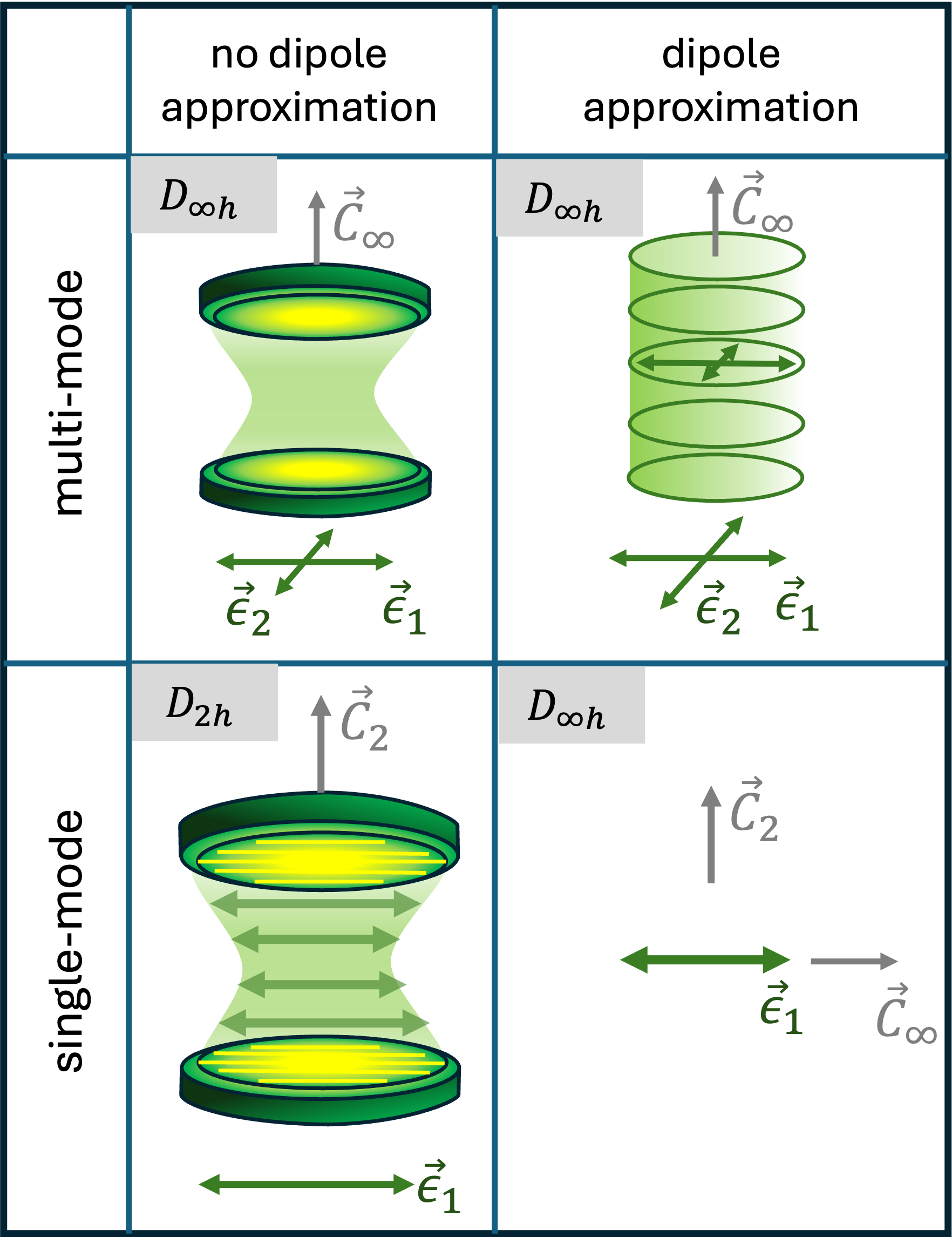}
    \caption{Cavity symmetry with and without the dipole approximation, respectively, in a multi-mode vs. single-mode approximation. 
             Polarization vectors are shown bidirectional to resemble the correct symmetry of the system.}
    \label{fig:cavity-sym}
\end{figure}

Our objective is to model a Fabry-Pérot cavity.\cite{yokoyama} 
Experimentally, such a cavity is formed by two mirrors, with the separation between them determining the frequencies $\omega_\alpha$ of the EM modes supported within the cavity. 
  First we discuss the bare cavity Hamiltonian (\ref{eq:Hph}) in absence of the matter system, i.e., 
  \begin{equation}
    \hat{H}_\text{bare} = \sum_\alpha \omega_\alpha \hat{\alpha}^\dagger \hat{\alpha} \;.
    \label{eq:cav_bare}
  \end{equation}
  The EM mode structure is altered in the presence of the matter system. 
The orientation of the cavity is defined by the wave vectors $\boldsymbol{k}_\alpha$, pointing in the direction of the mirrors.  
The wave vectors $\boldsymbol{k}_\alpha$ and frequencies $\omega_\alpha$ are related by $|\boldsymbol{k}_\alpha| = c\omega_\alpha$, where $c$ is the speed of light. 
For the bare cavity (Eq. \eqref{eq:cav_bare}, see also upper left corner of Fig. (\ref{fig:cavity-sym})), the symmetry corresponds to $D_{\infty h}$. 
Hence, the cavity Hamiltonian $\hat{H}_\text{bare}$ transforms symmetrically under symmetry operations within the $D_{\infty h}$ group. 
More concretely, $\hat{H}_\text{bare}$ transforms symmetrically under a $C_\infty$ rotation with respect to the principal axis of the cavity, under $C_2$ rotations perpendicular to it, under vertical reflections $\sigma_v$ containing the $C_\infty$-axis and under a horizontal reflection $\sigma_h$ normal to the $C_\infty$-axis, etc.

To maintain a $C_\infty$ rotation symmetry of the cavity Hamiltonian, for each mode $\alpha$ a perpendicularly polarized mode $\beta$ must exist. 
These two modes $\alpha$ and $\beta$ have the same frequency and same wave vector ($\boldsymbol{k}_\alpha = \boldsymbol{k}_\beta $) but have perpendicular field polarization such that $\boldsymbol{\epsilon}_\alpha \cdot \boldsymbol{\epsilon}_\beta = 0$.
This motivates a new notation where for each mode $\alpha$ with a field polarization of $\boldsymbol{\epsilon}_\alpha$, we introduce a corresponding mode $\bar{\alpha}$, with perpendicular field polarization $\bar{\boldsymbol{\epsilon}}_{\alpha}$, where $\boldsymbol{\epsilon}_\alpha \cdot \bar{\boldsymbol{\epsilon}}_{\alpha} = 0 $.
The bare cavity Hamiltonian (\ref{eq:cav_bare}) can then be rewritten as 
\begin{equation}
  \hat{H}_\text{bare} = \sum_\alpha^{N_\text{cav}/2} \omega_{\alpha} (\hat{\alpha}^\dagger \hat{\alpha} + \hat{\bar{\alpha}}^\dagger \hat{\bar{\alpha}} ) \;.
   \label{eq:Hph_sym} 
\end{equation}
To show that the Hamiltonian is rotationally invariant, we introduce the associated operators for a rotation with respect to the principal axis of the cavity around an arbitrary angle $\theta$
\begin{equation}
  \hat{C}_\theta  = \text{exp} \Big[ \theta \sum_\alpha^{N_\text{cav}/2} ( \hat{\alpha}^\dagger \hat{\bar{\alpha}} -\hat{\bar{\alpha}}^\dagger \hat{\alpha} ) \Big]   \; .
   \label{}
\end{equation}
With this operator it can easily be verified that the cavity Hamiltonian $\hat{H}_\text{cav}$ in the form (\ref{eq:Hph_sym}) is invariant under $C_\infty$ rotations. Thereby, the elementary operators transform as
\begin{equation}
   \begin{split}
       \hat{\alpha}^\prime & =\hat{C}_\theta^\dagger \hat{\alpha} \hat{C}_\theta  = \hat{\alpha}  \cos\theta + \hat{\bar{\alpha}} \sin\theta\\
       \hat{\bar{\alpha}}^\prime & =\hat{C}_\theta^\dagger \hat{\bar{\alpha}} \hat{C}_\theta  = \hat{\bar{\alpha}}  \cos\theta - \hat{\alpha} \sin\theta
   \end{split}
   \label{}
\end{equation}
and analogously for the creation operator $\hat{\alpha}^\dagger$. Hence, 
$\hat{C}_\theta^\dagger \hat{H}_\text{bare}\hat{C}_\theta = \hat{H}_\text{bare}$. 
Also, the Hamiltonian in the form (\ref{eq:Hph_sym}) readily transforms symmetrically under $\sigma_v$ reflections (or $C_{2}$ rotations ; $\hat{C}_{2}(\boldsymbol{n}) = \hat{\sigma}_{v}(\boldsymbol{n})$)
\begin{equation}
  \begin{split}
  \hat{\sigma}_{v}(\boldsymbol{n})  = \text{exp} \Big[ & i \pi \sum_\alpha^{N_\text{cav}} 
  \Big(
  n_2^2
  \hat{\alpha}^\dagger \hat{\alpha} + n_1^2 \hat{\bar{\alpha}}^\dagger \hat{\bar{\alpha}} \\
  & - n_1 n_2 ( \hat{\bar{\alpha}}^\dagger \hat{\alpha} + \hat{\alpha}^\dagger \hat{\bar{\alpha}} ) \Big)\Big]   \; .
  \end{split}
   \label{}
\end{equation}
The two dimensional vector $\boldsymbol{n} = (n_1 , n_2)^T$ is a unit vector in the $\boldsymbol{\epsilon}_\alpha$, $\bar{\boldsymbol{\epsilon}}_\alpha$ plane and contains the reflection plane, or, respectively, the rotation axis.
The elementary operators transform as
\begin{equation}
   \begin{split}
     \hat{\alpha}^\prime & =\hat{\sigma}_{v}^\dagger \hat{\alpha} \hat{\sigma}_{v}  = (1 - 2 n_2^2 ) \hat{\alpha} + 2 n_1 n_2 \hat{\bar{\alpha}}  \\
     \hat{\bar{\alpha}}^\prime & =\hat{\sigma}_{v}^\dagger \hat{\bar{\alpha}} \hat{\sigma}_{v}  = (1 - 2 n_1^2 ) \hat{\bar{\alpha}} + 2 n_1 n_2 \hat{\alpha}  \\
   \end{split}
   \label{}
\end{equation}
The horizontal reflection $\sigma_h$, is defined as a reflection normal to the $\boldsymbol{k}_\alpha$ vectors where the reflection plane is located in a surface spanned by $\boldsymbol{\epsilon}$ and $\bar{\boldsymbol{\epsilon}}$.
In the dipole approximation, the electric field is homogeneous and oriented along the polarization vectors, so the horizontal reflection leaves the electric field invariant and therefore
\begin{equation}
  \hat{\sigma}_h = 1\;.
  \label{}
\end{equation}
This, however, only holds in the dipole approximation, while for example a circularly polarized cavity changes the polarization under a $\hat{\sigma}_h$ operation as has been shown by Riso et al.\cite{riso2023strong}. 
As usual, higher symmetry operations as an inversion $i$ or $S_n$-rotation can be represented by sequentially applying the symmetry operations above.

\subsection{Point-group symmetry for the light-matter system}
To maintain the symmetries for the coupled light-matter system, also the displacement field should be set up in a basis of perpendicularly polarized modes
\begin{equation}
\hat{\boldsymbol{D}} = \sum_\alpha^{N_\text{cav}/2} \sqrt{\frac{\lambda_\alpha^2 \omega_\alpha}{2}} \Big[ \boldsymbol{\epsilon}_\alpha (\hat{\alpha}^\dagger + \hat{\alpha}) +  \bar{\boldsymbol{\epsilon}}_\alpha (\hat{\bar{\alpha}}^\dagger + \hat{\bar{\alpha}} ) ) \Big]\;.
   \label{eq:bil_sym}
\end{equation}
Note that to retain the symmetries of the cavity, the coupling strength along two perpendicular polarized modes must be equal $\lambda_\alpha = \bar{\lambda}_\alpha  $.
As the orientations of $\boldsymbol{\epsilon}_\alpha$ and $\bar{\boldsymbol{\epsilon}}_\alpha$ are arbitrary as long as $\boldsymbol{k}_\alpha$, $\boldsymbol{\epsilon}_\alpha$ and $\bar{\boldsymbol{\epsilon}}_\alpha$ are orthonormal, a linear combination of the polarization vectors can be formed. 
The displacement field operator might therefore also be written in terms of two unified field polarizations $\boldsymbol{\epsilon}\equiv\boldsymbol{\epsilon}_\alpha$ and $\bar{\boldsymbol{\epsilon}}\equiv\bar{\boldsymbol{\epsilon}}_\alpha$ for all $\alpha$, so that 
\begin{equation}
\hat{\boldsymbol{D}} = \sum_\alpha^{N_\text{cav}/2} \sqrt{\frac{\lambda_\alpha^2 \omega_\alpha}{2}} \Big[ \boldsymbol{\epsilon} (\hat{\alpha}^\dagger + \hat{\alpha}) +  \bar{\boldsymbol{\epsilon}} (\hat{\bar{\alpha}}^\dagger + \hat{\bar{\alpha}} ) ) \Big].
   \label{eq:efield_displace}
\end{equation}
Note that the displacement operator $\hat{\boldsymbol{D}}$ is not invariant under $C_\infty$ rotations. 
The rotations can be written as 
\begin{equation}
   \begin{split}
   \hat{C}_\theta^\dagger \hat{\boldsymbol{D}} \hat{C}_\theta & =  \sum_\alpha \sqrt{\frac{\lambda_\alpha^2 \omega_\alpha}{2}} \Big[ \boldsymbol{\epsilon}^\prime (\hat{\alpha}^\dagger + \hat{\alpha}) +  \bar{\boldsymbol{\epsilon}}^\prime (\hat{\bar{\alpha}}^\dagger + \hat{\bar{\alpha}} ) ) \Big], \\
   \end{split}
   \label{}
\end{equation}
with the new basis 
\begin{equation}
   \begin{split}
     \boldsymbol{\epsilon}^\prime = (\cos\theta \boldsymbol{\epsilon} - \sin\theta \bar{\boldsymbol{\epsilon}}) \\
     \bar{\boldsymbol{\epsilon}}^\prime = (\cos\theta  \bar{\boldsymbol{\epsilon}} + \sin\theta  \boldsymbol{\epsilon} )
   \end{split}
   \label{},
\end{equation}
i.e., the polarization vectors change. For example, for a rotation of $\theta=\pi$, the displacement field is inverted: $ \hat{C}_\pi^\dagger \hat{\boldsymbol{D}} \hat{C}_\pi = - \hat{\boldsymbol{D}}$.
\\
However, the bilinear coupling $\boldsymbol{D}\cdot\boldsymbol{d}$ is invariant under a $\hat{C}_\theta$ rotation:
\begin{equation}
   \hat{C}_\theta^\dagger \hat{\boldsymbol{D}}\cdot\hat{\boldsymbol{d}} \hat{C}_\theta =  \hat{\boldsymbol{D}}\cdot \hat{\boldsymbol{d}} .
   \label{}
\end{equation}
This is ensured by the fact that the  $\boldsymbol{\epsilon}\cdot\hat{\tilde{\boldsymbol{d}}}$ and $\bar{\boldsymbol{\epsilon}}\cdot\hat{\tilde{\boldsymbol{d}}}$ components of the molecular dipole operator transform in the same manner as the displacement field $\hat{\boldsymbol{D}}$.
In the example above, with $\theta = \pi$, the  $\boldsymbol{\epsilon}\cdot\hat{\tilde{\boldsymbol{d}}}$ and $\bar{\boldsymbol{\epsilon}}\cdot\hat{\tilde{\boldsymbol{d}}}$ components are inverted as well.\footnote{For electronic operators, the transformation can be found in the literature, see, for example, Refs.  \citenum{bishop1993group} and \citenum{helgaker2013molecular}. }
A similar behavior is found for the vertical $C_{2}$ rotations and $\sigma_v$ reflections. 
This holds for all symmetry operations of the group.

For the combined light-matter system, due to the parametrically introduced nuclei, the $D_{\infty h}$ symmetry of the cavity may be broken. 
For the polaritonic wave function $\ket{\Psi}$, in general, 
\begin{equation}
      \braket{\Psi| \hat{\alpha}^\dagger \hat{\alpha} |\Psi} \ne \braket{\Psi| \hat{\bar{\alpha}}^\dagger \hat{\bar{\alpha}} |\Psi}
   \label{eq:symcav}, 
\end{equation}
i.e., the two displacement modes may exhibit differing levels of excitation.
Exceptions are atoms and linear molecules oriented parallel along $\boldsymbol{k}$. 
In all systems, however, the energy is invariant among rotations of the polarization vectors.\footnote{Note that the same is not true for a molecule in a linearly polarized cavity.} 

We note that in our description the cavity is neither linearly nor circularly polarized and that also complex valued polarization vectors can be used.
A unitary transformation which shows the equivalence can be found in Appendix \ref{sec:circ}.

\subsection{Single-mode approximation}
For a numerical treatment, the number of modes to be included is typically limited. 
In the (physical) single-mode approximation, there are two modes  of the same frequency and same coupling strength but two perpendicular polarizations ($N_\text{cav}=2$). 
Note that the term single-mode approximation is somewhat ambiguous. In the physical sense it means that the EM field is described only by a single frequency.  
This approximation is motivated by the fact that only a small energy window is considered for the light-matter interactions where electronic transitions are energetically close to one given cavity frequency. 
However, one frequency of the EM field is still composed of two orthogonal modes with perpendicular field polarizations. 
To obtain a single-mode approximation in a mathematical sense, therefore, for a linearly polarized cavity, it is sufficient to let $\bar{\lambda}_\alpha\rightarrow 0 $ as one of two field polarizations is not coupled to the matter system (in addition to $N_\text{cav}=2$).
For an unpolarized cavity in the single-mode approximation, the displacement field can be written in compact notation as
\begin{equation}
\hat{\boldsymbol{D}} = \sqrt{\frac{\lambda^2 \omega}{2}} \Big[ \boldsymbol{\epsilon} (\hat{\alpha}^\dagger + \hat{\alpha}) +  \bar{\boldsymbol{\epsilon}} (\hat{\bar{\alpha}}^\dagger + \hat{\bar{\alpha}} ) ) \Big]
   \label{}
   .
\end{equation}
As shown in Fig. \ref{fig:cavity-sym}, depending on the approximation in the Hamiltonian, the cavity symmetry is different. 
In the upper left corner, a general Fabry-Pérot cavity is depicted which is characterized by the principal rotation axis $C_\infty$  pointing in the direction of the two mirrors. The system has an overall $D_{\infty h}$ symmetry. 
For the polarized case, the symmetry is reduced as for example for linearly polarized light with polarization direction $\boldsymbol{\epsilon}$ - see lower left panel in Fig. \ref{fig:cavity-sym}. 
The principal rotation axis is then reduced from $C_\infty$ to $C_2$, noting that two orthogonal $C_2$ rotations remain.

When applying the dipole approximation it is assumed that the wavelength of the EM field is large over the dimension of the molecule and that the matter-system is not close to the boundaries. 
In Fig. \ref{fig:cavity-sym} this is realized by removing the mirrors so that the electric field has no boundaries. 
This can be understood as an rotationally symmetric electric field pointing in an arbitrary direction in the $xy$-plane - see upper right corner of Fig. \ref{fig:cavity-sym}. 
When additionally assuming that one field polarization is not coupled to the molecule, one of the polarization vectors is removed by which the electric field is given a predefined polarization. 
This can be seen in Fig. \ref{fig:cavity-sym} in the lower-right corner, where a homogeneous electric field is oriented along the $\epsilon_x$-vector.  
In consequence, an additional $C_\infty$ rotation axis is formed in the direction of the polarization vector. 
This means that the symmetry of the system in the dipole approximation together with the single-mode treatment is higher ($D_{\infty h}$) than in the setup for a linearly polarized cavity ($D_{2h}$), i.e., the lower left part of Fig. \ref{fig:cavity-sym}. 
Also, when comparing the results of an unpolarized Fabry-Pérot cavity to a linearly polarized cavity, the point-group might be the same ($D_{\infty h}$), but the principal axis of the system is shifted by $90^\circ$, which leads to a different symmetry. 

\subsection{Unpolarized Polaritonic CC Theory}
The polaritonic CC wave function ansatz is\cite{haugland2020coupled}
\begin{equation}
   \ket{\Psi_\text{CC}}=  \text{e}^{\hat{Q}} \ket{0,0}
   \label{}
\end{equation}
with the cluster operator $\hat{Q}$.
The state $\ket{0,0}$ is the Fermi-vacuum and represents the HF determinant with the photonic vacuum.
Both numbers in $\ket{0,0}$ can be understood as occupation number vectors (ONVs), where the first number designates the electronic ONV  and the second the photonic ONV. 
For the electronic space, a ``$0$'' designates the case where all electrons are located in occupied orbitals, the state $\ket{\substack{a\\i},0}$ is a determinant where an electron is excited from the orbital $i$ into the virtual orbital $a$ and similarly for double excitations $\ket{\substack{a b\\ i j},0}$.
In the photonic space a ``$0$'' represents the absence of photons in all modes, therefore the physical vacuum.
The state $\ket{0,1_n}$ designates a state where one photon is in the $n$'th mode, and the state $\ket{0,1_n 1_{\bar{n}}}$ are two photons, one in mode $n$ and the other in the perpendicular polarized mode $\bar{n}$.
This scheme can easily be generalized to include higher photonic excitations and other modes.

In this work the CCSD-12-SD truncation scheme is employed, with the cluster operator truncated as\cite{haugland2020coupled}
\begin{equation}
   \hat{Q} = \hat{T}_1 + \hat{T}_2 +\hat{S}_1^1 +\hat{S}_2^1 +\hat{\Gamma}_1 + \hat{\Gamma}_2\;.
   \label{}
\end{equation}
Here, $\hat{T}_1$ and $\hat{T}_2$ are the standard electronic single and double excitation operators, $\hat{S}_1^1$ and $\hat{S}_2^1$ are the mixed operators that include one-photon creation and $\hat{\Gamma}_1$ and $\hat{\Gamma}_2$ are the bare on-e and two-photon creation operators.
The $\hat \Gamma_1$ operator is set up as
\begin{equation}
   \hat \Gamma_1 = \sum_\alpha \left( \gamma^\alpha \hat{\alpha}^\dagger + \gamma^{\bar{\alpha}} \hat{\bar{\alpha}}^\dagger
   \right)
   \label{eq:G1}
\end{equation}
with the two sets of amplitudes corresponding to perpendicular field polarizations ($\gamma^\alpha$ and $\gamma^{\bar{\alpha}}$).
In the same manner the $\hat{\Gamma}^2$ operator can be set up as 
\begin{equation}
   \begin{split}
      \hat{\Gamma}_2 & = \sum_{\alpha<\beta} 
      \left(
      \gamma^{\alpha \beta} \hat{\alpha}^\dagger \hat{\beta}^\dagger +  \gamma^{\bar{\alpha} \bar{\beta}} \hat{\bar{\alpha}}^\dagger \hat{\bar{\beta}}^\dagger + \gamma^{\alpha \bar{\beta}} \hat{\alpha}^\dagger \hat{\bar{\beta}}^\dagger + \gamma^{\bar{\alpha} \beta} \hat{\bar{\alpha}}^\dagger \hat{\beta}^\dagger
      \right)
   \end{split}
   \label{eq:G2}
\end{equation}
Regarding  the $\hat\Gamma_2$ operator, 
when introducing unrestricted sums, the diagonal elements of $\gamma^{\alpha\alpha}$ and $\bar{\bar{\gamma}}^{\alpha\alpha}$ would falsely be scaled by a factor of $\frac{1}{2}$.
This must be taken into account by using 
\begin{equation}
   \gamma^{\alpha\beta} \rightarrow (1 + \delta_{\alpha\beta}) \gamma^{\alpha\beta}
   \label{eq:amplitude_G2}
\end{equation}
and similarly for $\gamma^{\bar{\alpha}\bar{\beta}}$, but not for $\gamma^{\alpha\bar{\beta}}$ as the latter corresponds to an off-diagonal block and in general is not symmetric.
When further exploiting the symmetry $\gamma^{\alpha\bar{\beta}} = \gamma^{\bar{\beta}\alpha}$ and the commutator $[\alpha^\dagger,\bar{\beta}^\dagger]=0$, the $\hat \Gamma_2$ operator becomes
\begin{equation}
   \begin{split}
      \hat{\Gamma}_2 & = \frac{1}{2} \sum_{\alpha\beta} 
      \left(
      \gamma^{\alpha \beta} \hat{\alpha}^\dagger \hat{\beta}^\dagger +  \gamma^{\bar{\alpha} \bar{\beta}} \hat{\bar{\alpha}}^\dagger \hat{\bar{\beta}}^\dagger + 2 \gamma^{\bar{\alpha} \beta} \hat{\alpha}^\dagger \hat{\bar{\beta}}^\dagger \right) \;.\\
   \end{split}
   \label{eq:G2-2}
\end{equation}
The exploitation of polarization symmetry works therefore completely analogously to the exploitation of spin symmetry.
E.g., in a spin-unrestricted treatment, the $\hat T_2$ operator is parameterized as 
\begin{equation}
   \begin{split}
      \hat{T}_2  & = \frac{1}{4} \sum_{abij}  t_{abij} \;  \hat{a}^\dagger  \hat{i}\hat{b}^\dagger \hat{j}  + \frac{1}{4}\sum_{aibj} t_{\bar{a}\bar{b}\bar{i}\bar{j}} \; \hat{\bar{a}}^\dagger  \hat{\bar{i}}\hat{\bar{b}}^\dagger \hat{\bar{j}}  \\
      & +  \sum_{aibj} t_{a\bar{b}i\bar{j}}  \hat{a}^\dagger \hat{i} \hat{\bar{b}}^\dagger \hat{\bar{j}}\;.
   \end{split}
   \label{eq:symS2}
\end{equation}
where for the amplitudes we used that $t_{ai\bar{b}\bar{j}} = - t_{a\bar{i}\bar{b}j} = - t_{\bar{a}ib\bar{j}} = t_{\bar{a}\bar{i}bj}$.
This partitioning for the photonic and electronic indices can equally be applied for the $\hat{S}_1^1$ and  $\hat{S}_2^1$ operators:
\begin{equation}
   \begin{split}
      \hat S_1^1 & = \sum_{ai \alpha} (s_{ai}^\alpha  \hat{\alpha}^\dagger + s_{ai}^{\bar{\alpha}}  \hat{\bar{\alpha}}^\dagger ) \hat{a}^\dagger \hat{i} \\
      & + \sum_{ai \alpha} (s_{\bar{a}\bar{i}}^\alpha  \hat{\alpha}^\dagger + s_{\bar{a}\bar{i}}^{\bar{\alpha}}  \hat{\bar{\alpha}}^\dagger ) \hat{\bar{a}}^\dagger \hat{\bar{i}} \\
   \end{split}
   \label{eq:symS1}
\end{equation}
\begin{equation}
   \begin{split}
   \hat S_2^1  & = \frac{1}{4}\sum_{abij \alpha} ( s_{aibj}^\alpha  \hat{\alpha}^\dagger + s_{aibj}^{\bar{\alpha}}  \hat{\bar{\alpha}}^\dagger) \hat{a}^\dagger  \hat{i}\hat{b}^\dagger \hat{j} \\
   & + \frac{1}{4}\sum_{abij \alpha} ( s_{\bar{a}\bar{i}\bar{b}\bar{j}}^\alpha  \hat{\alpha}^\dagger + s_{\bar{a}\bar{i}\bar{b}\bar{j}}^{\bar{\alpha}}  \hat{\bar{\alpha}}^\dagger) \hat{\bar{a}}^\dagger  \hat{\bar{i}}\hat{\bar{b}}^\dagger \hat{\bar{j}} \\
   & + \sum_{abij \alpha} ( s_{ai\bar{b}\bar{j}}^\alpha  \hat{\alpha}^\dagger + s_{ai\bar{b}\bar{j}}^{\bar{\alpha}}  \hat{\bar{\alpha}}^\dagger) \hat{a}^\dagger  \hat{i}\hat{\bar{b}}^\dagger \hat{\bar{j}}. 
   \end{split}
   \label{eq:symS2}
\end{equation}
If the molecule is symmetric for rotations along the orientation of the cavity, the symmetry $\gamma^\alpha = \gamma^{\bar{\alpha}}$,  $s^\alpha_{ai} = s^{\bar{\alpha}}_{ai}$ and $s^\alpha_{abij} = s^{\bar{\alpha}}_{abij}$ can be employed to give an even more compact notation.
E.g., the $\hat \Gamma_1$ operator would be set up as
\begin{equation}
  \hat \Gamma_1 = \sum_\alpha \gamma^\alpha ( \hat{\alpha}^\dagger + \hat{\bar{\alpha}}^\dagger ) 
   \label{eq:symG1}
\end{equation}
and similarly for $\hat S_1^1$ and $\hat S_2^1$.
The CC amplitudes are solved in a non-linear set of projected equations 
\begin{equation}
   \braket{\mu,\nu| \hat{\tilde{H}} | 0,0}  = 0
   \label{}
\end{equation}
with $\mu$ designating an electronically excited determinant and $\nu$ an excitation of the EM field.
The similarity-transformed Hamiltonian  $\hat{\tilde{H}}$ is given as
\begin{equation}
  \hat{\tilde{H}} = \text{e}^{-\hat{Q}} \hat{H} \text{e}^{\hat{Q}}\;.
   \label{eq:simham}
\end{equation}
A detailed description of how the CC amplitude equations are solved can be found in Ref. \citenum{monzel2024diagram}.
\\
To calculate properties and densities at the CC level of theory, the left-hand side solution of the similarity-transformed Hamiltonian matrix is required.
The left hand side wave function is given as\cite{castagnola2024polaritonic,monzel2024diagram}
\begin{equation}
  \bra{\Psi_\text{CC}^\text{L}} = \bra{0,0} ( 1 + \hat{\Lambda} ) \text{e}^{-\hat{Q}}
  \label{eq:lambda_equation}
\end{equation}
with $\hat{\Lambda}$ being a de-excitation operator truncated in the same scheme as $\hat{Q}$.
The coefficients in $\hat{\Lambda}$ can be solved by a set of projected equations
\begin{equation}
  \braket{0,0|(1 + \hat{\Lambda}) \hat{\tilde{H}} |\mu,\nu } = 0\;.
  \label{}
\end{equation}
With a converged set of $\Lambda$-amplitudes, polaritonic expectation values can be obtained in the familiar manner via
\begin{equation}
  \braket{O} = \braket{\Psi_\text{CC}^L| \hat{O} |\Psi_\text{CC}}\;.
  \label{}
\end{equation}
This also allows for the calculation of the electron and photon density matrices.
E.g., the one-electron density matrix is obtained as
\begin{equation}
  \gamma_{pq} = \braket{\Psi_\text{CC}^L| \hat{p}^\dagger \hat{q} |\Psi_\text{CC}}\;.
  \label{}
\end{equation}
A detailed description of how the $\Lambda$-equations are solved and how the density matrices are calculated can be found in Ref. \citenum{monzel2024diagram}.
The one-electron density matrix is used with the molecular orbitals $\phi_p(\boldsymbol{r})$ to assemble the one-electron density
\begin{equation}
  \rho(\boldsymbol{r}) = \sum_{pq} \gamma_{pq} \phi_p^*(\boldsymbol{r}) \phi_q(\boldsymbol{r}).
  \label{eq:one-el-density}
\end{equation}
\\
Excited states can be obtained from the converged set of CC-amplitudes via the equation-of-motion-CC (EOM-CC) parametrization\cite{stanton1993eom}
\begin{equation}
  \hat{\tilde{H}} \hat{R} \ket{0,0} = E_\text{exc} \hat{R} \ket{0,0},
  \label{eq:eom-cc}
\end{equation}
where the excitation operator $\hat{R}$ is truncated in the same scheme as the cluster operator $\hat{Q}$. 
Like in electronic CC theory, the excitation operator $\hat{R}$ commutes with the cluster operator $\hat{Q}$.
The coefficients in $\hat{R}$ can be obtained by a Davidson-like algorithm.\cite{davidson}
But, while the cluster operator $\hat{Q}$ and the de-excitation operator $\hat{\Lambda}$  transform in a totally symmetric manner, the excitation operator $\hat{R}$ transforms according to the irreducible representation of the excitation. 
A detailed description of how the EOM-CC equations are solved for the coefficients in $\hat{R}$ can be found in Ref. \citenum{monzel2024diagram}.
\\

\subsection{Exploiting Point-Group Symmetry in Cavity-QED Calculations}
\label{sec:symmetry}
Exploiting point-group symmetry in the context of direct-product decomposition in quantum-chemical calculations leads to significant speedups in computational time\cite{stanton1991symmetry} and enables the targeted calculation of states of different irreducible representations.
For the treatment of electronic operators in terms of the direct-product decomposition, the reader is referred to Ref.\citenum{stanton1991symmetry}.
The following discussion will focus on the bilinear coupling and the photonic Hamiltonian. 

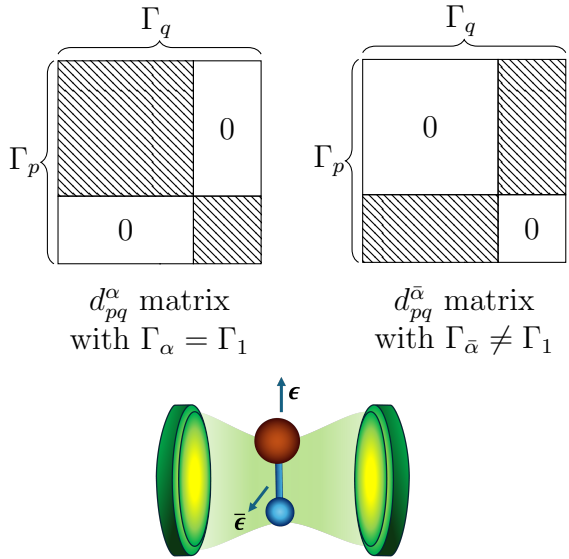
\begin{figure}[t]
\centering
\begin{tikzpicture}[scale=0.9]
   \draw [pattern=north west lines] (2,0)   rectangle (3,1  )  ;
   \draw [pattern=north west lines] (0,1)   rectangle (2,3  )  ;
   \draw [] (0,0)   rectangle (2,1  ) node[midway] {0};
   \draw [] (2,1)   rectangle (3,3  ) node[midway] {0};

  \draw [decorate, decoration={brace,amplitude=5pt}, xshift=+4pt, yshift=0pt] (-0.2,0) -- (-0.2,3)    node[midway,xshift=-10pt,text width=0.5cm] {$\Gamma_p$ };
  \draw [decorate, decoration={brace,amplitude=5pt,raise=1pt}, yshift=-4pt] ( 0.0,3.2) -- ( 3.0,3.2)    node[midway,yshift=13pt,text width=0.5cm] {$\Gamma_q$ };

   \draw [] (1.5,0) -- node[below,yshift=-5pt,midway]{$d_{pq}^\alpha$ matrix }(1.5,0)  ;
   \draw [] (1.5,0) -- node[below,yshift=-20pt,midway]{with $\Gamma_\alpha = \Gamma_1$}(1.5,0)  ;

\end{tikzpicture}
\quad
\begin{tikzpicture}[scale=0.9]
   \draw [] (2,0)   rectangle (3,1  ) node[midway] {0} ;
   \draw [] (0,1)   rectangle (2,3  ) node[midway] {0} ;
   \draw [pattern=north west lines] (0,0)   rectangle (2,1  ) ;
   \draw [pattern=north west lines] (2,1)   rectangle (3,3  ) ;

  \draw [decorate, decoration={brace,amplitude=5pt}, xshift=+4pt, yshift=0pt] (-0.2,0) -- (-0.2,3)    node[midway,xshift=-10pt,text width=0.5cm] {$\Gamma_p$ };
  \draw [decorate, decoration={brace,amplitude=5pt,raise=1pt}, yshift=-4pt] ( 0.0,3.2) -- ( 3.0,3.2)    node[midway,yshift=13pt,text width=0.5cm] {$\Gamma_q$ };

   \draw [] (1.5,0) -- node[below,yshift=-5pt,midway]{$d_{pq}^{\bar{\alpha}}$ matrix }(1.5,0)  ;
   \draw [] (1.5,0) -- node[below,yshift=-20pt,midway]{with $\Gamma_{\bar{\alpha}} \ne \Gamma_1$}(1.5,0)  ;

\end{tikzpicture}\\
\includegraphics[width=0.4\linewidth]{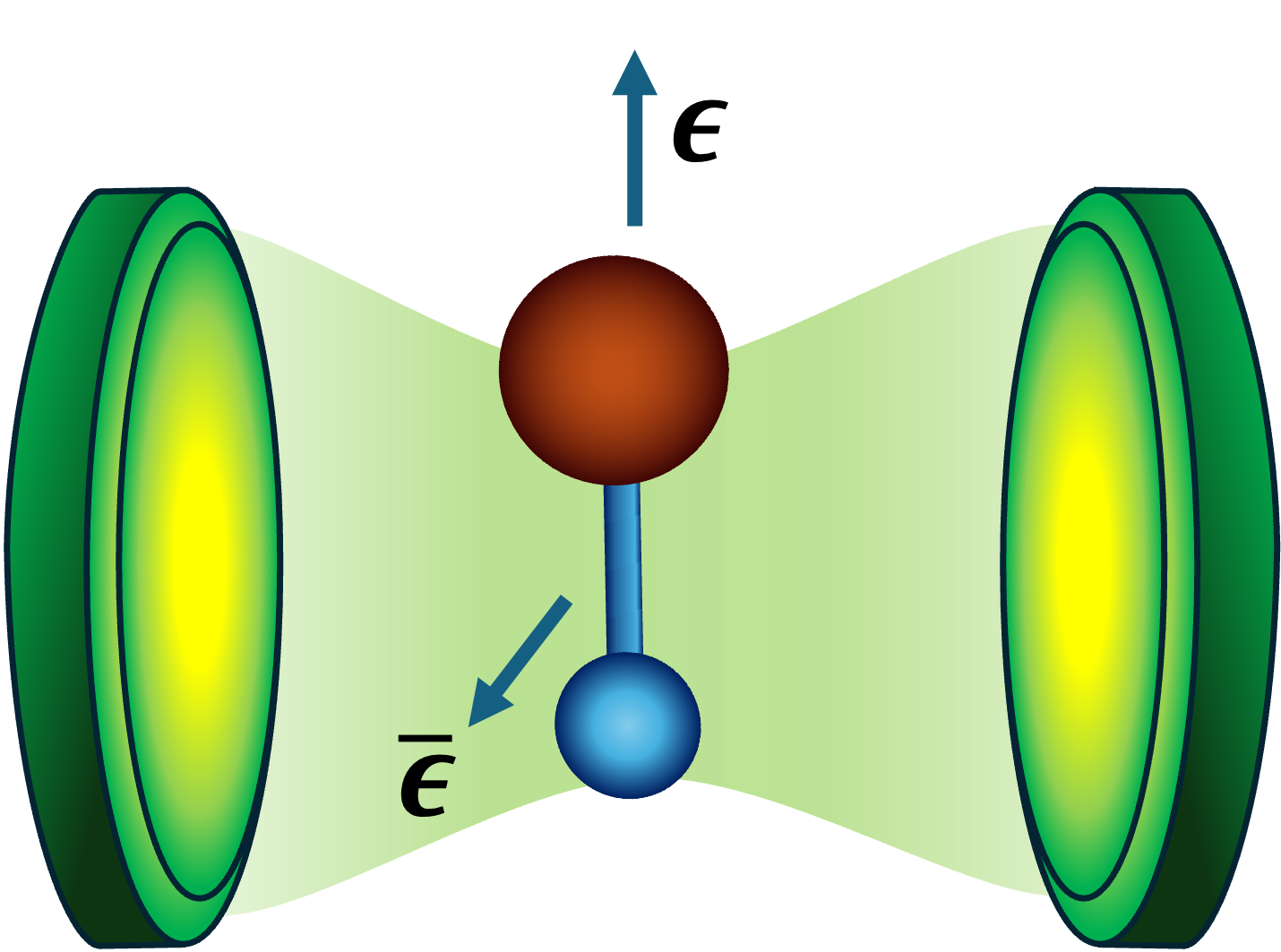}\\
\caption{Exemplary matrix representation of the bilinear coupling $\tilde{d}_{pq}^\alpha$ and $\tilde{d}_{pq}^{\bar{\alpha}}$. 
         The polarization $\boldsymbol{\epsilon}$ is here equal to the totally symmetric irreducible representation $\Gamma_\alpha = \Gamma_1$ and the electronic indices become block diagonal.
         The second polarization $\bar{\boldsymbol{\epsilon}}$ is not totally symmetric and the occupation for the electronic indices is on the off-diagonal blocks.}
\label{fig:matrix}
\end{figure}

As discussed before, the Hamiltonian is totally symmetric for symmetry operations within the point group. 
Consequently, all operators in the Hamiltonian are characterized by the totally symmetric irreducible representation $\Gamma_1$.\cite{bishop1993group}
For the cavity Hamiltonian $\hat{H}_\mathrm{cav}$ this property is rather obvious, as by construction it is build with the number operator $\hat{\alpha}^\dagger\hat{\alpha}$ and the direct product of an irreducible representation with itself always contains the totally symmetric representation
\begin{equation}
  \Gamma_1 \in \Gamma_\alpha \otimes \Gamma_\alpha\;.
  \label{eq:irrep_Hph}
\end{equation}
The bilinear coupling operator in second quantization for two perpendicular polarized modes can be written in the form 
\begin{equation}
	\begin{split}
  \hat{H}_\text{bil} & = \sum_\alpha^{N_\text{cav}/2} \sum_{pq} \tilde{d}_{pq}^{\alpha}  (\hat{\alpha} + \hat{\alpha}^\dagger) \hat{p}^\dagger \hat{q} \\
      &+ \sum_\alpha^{N_\text{cav}/2} \sum_{pq} \tilde{d}_{pq}^{\bar{\alpha}}  (\hat{\bar{\alpha}} + \hat{\bar{\alpha}}^\dagger) \hat{p}^\dagger \hat{q} \;.
      \label{}
	\end{split}
   \label{eq:bil}
\end{equation}
A single term of the bilinear coupling operator is totally symmetric if the direct product of the irreducible representation of the elementary operators contains the $\Gamma_1$ representation, therefore
\begin{equation}
  \Gamma_1 \in \Gamma_p \otimes \Gamma_q \otimes \Gamma_\alpha\;.
   \label{eq:bilirrep}
\end{equation}
Depending on the molecule and its orientation, the bilinear coupling can be grouped in the following way: 
If the molecule has a dipole moment oriented along one polarization vector, the photonic index is totally symmetric ($\Gamma_\alpha = \Gamma_1$) and the bilinear coupling integrals $\tilde{d}^\alpha_{pq}$ become block diagonal in the electronic indices ($\Gamma_p = \Gamma_q$).
On the other hand, if the molecule has no dipole moment, or a dipole moment not aligned with the polarization vector, the irreducible representation of the photonic index is not totally symmetric ($\Gamma_\alpha \ne \Gamma_1$) and hence the diagonal blocks for the electronic indices vanish ($\Gamma_p \ne \Gamma_q$).
Thus, the implementation needs to be flexible enough to handle both block-diagonal matrices and matrices with nonzero blocks on the off-diagonals - see Fig. \ref{fig:matrix}.

In principle, the orientation of the polarization vectors is arbitrary as long as they are perpendicular to $\mathbf k$ and among each other. 
Here, we chose one polarization vector, e.g. $\boldsymbol{\epsilon}$, to be aligned parallel to the  dipole moment of the molecule and the second polarization vector $\bar{\boldsymbol{\epsilon}}$ perpendicular to it. 
The bilinear coupling operator then reads: 
\begin{equation}
	\begin{split}
    \hat{H}_\text{bil} & = \sum_{\alpha \in \Gamma_1} \sum_{\Gamma_\text{el}}  \sum_{\substack{p \in \Gamma_\text{el} \\ q \in \Gamma_\text{el}}} \tilde{d}_{pq}^{\alpha}  (\hat{\alpha} + \hat{\alpha}^\dagger) \hat{p}^\dagger \hat{q} \\
    & + \sum_{\substack{\Gamma_1 \in \\ \Gamma_\text{ph}\otimes\Gamma_p \otimes \Gamma_q}} \sum_{\substack{p \in \Gamma_p \\ q \in \Gamma_q \\ \bar{\alpha} \in \Gamma_\text{ph} }} \tilde{d}_{pq}^{\bar{\alpha}} (\hat{\bar{\alpha}} + \hat{\bar{\alpha}}^\dagger) \hat{p}^\dagger \hat{q} \;,
      \label{}
	\end{split}
   \label{eq:bil}
\end{equation}
where the second term only runs over irreducible representations where $\Gamma_1 \in \Gamma_\text{ph} \otimes \Gamma_p \otimes \Gamma_q$ is fulfilled and $\Gamma_\text{ph}\neq \Gamma_1$.
An exemplary representation of the bilinear coupling operator in matrix representation can be found in Fig (\ref{fig:matrix}).

Handling of symmetry based on the direct product decomposition in the context of CC theory is well known \cite{stanton1991symmetry,Gauss1991coupled} and has been exploited in many quantum-chemical program packages. 
It ensures that the number of floating-point operations can be reduced by a factor of $h^2$, with  the order of the group,  $h$.\cite{Gauss1991coupled}
For use within QED-CC theory, the photonic indices must be taken into account. 

As the cluster operator $\hat{Q}$ must be totally symmetric, it is obvious that the $\gamma^\alpha$ amplitude vanishes when $\hat{\alpha}^\dagger$ is not totally symmetric itself.
In other words, the molecule must have a dipole moment along $\boldsymbol{\epsilon}$ in order to have non-vanishing contributions in $\gamma^\alpha$.
For the same reason in the mixed amplitudes  $s_{ia}^\alpha$ and $s_{ijab}^\alpha$ the electronic indices do not have to contain the totally symmetric irreducible representation as long as the full operator is totally symmetric.
An exemplary contraction from the QED-CC amplitude equations is 
\begin{equation}
   \begin{split}
     & \bra{\substack{a\\i},0}[[\hat{H}_\text{bil}, \hat{T}_2],\hat{S}_1^1]\ket{0,0}\leftarrow 
     \begin{minipage}{0.2\linewidth}\includegraphics[width=1.0\linewidth]{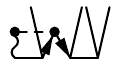}\end{minipage} = \\
     & = - P(ij)\sum_{\Gamma_\text{el}} \sum_{\substack{e \in \Gamma_\text{el} \\ m \in \Gamma_\text{el}}} \sum_{\beta\in\Gamma_1} s_{ei}^{\beta} \; t_{abmj} \; d_{me}^{\beta} \;\\
       & -P(ij)\sum_{\substack{\Gamma_1 \in \\ \Gamma_e \otimes \Gamma_m \otimes  \Gamma_{\text{ph}}  }} \sum_{\substack{e \in \Gamma_e \\ m \in \Gamma_m}} \sum_{\beta \in \Gamma_{\text{ph}}} s_{ei}^{\bar{\beta}} \; t_{abmj} \; d_{me}^{\bar{\beta}}\;,
   \end{split}
\end{equation}
where the polarization vector $\boldsymbol{\epsilon}$ is parallel to the molecular dipole moment and the second polarization vector $\bar{\boldsymbol{\epsilon}}$ perpendicular to it. 
The sum over $\Gamma_\text{el}$ is over all irreducible representations in the group and further $\Gamma_\text{ph}\neq\Gamma_1$. 
The amplitudes $s_{ei}^\alpha$ and $s_{ei}^{\bar{\alpha}}$ can be grouped into blocks similar to the bilinear coupling integrals $d_{pq}^\alpha$ and $d_{pq}^{\bar{\alpha}}$, see Fig. \ref{fig:matrix}.

Exploiting the block-structure of tensors in the above manner ensures that only the potentially non-zero elements are computed which leads to time and memory savings.
Further, it allows to make use of the usual integral-packing strategies for the permutations of indices of the different irreducible representations.
For example the mixed amplitudes $s_{aibj}^{\alpha}$ can be packed exploiting the block structure 
\begin{equation}
   \begin{split}
  & s^{\Gamma_\text{ph}}_{\Gamma_A \Gamma_B \Gamma_A \Gamma_C} = -s^{\Gamma_\text{ph}}_{\Gamma_A \Gamma_B \Gamma_C \Gamma_A } \\
 =& s^{\Gamma_\text{ph}}_{ \Gamma_B \Gamma_A \Gamma_C \Gamma_A}= -s^{\Gamma_\text{ph}}_{ \Gamma_B \Gamma_A \Gamma_A \Gamma_C} \;.
   \end{split}
   \label{}
\end{equation}
Similar to regular CC theory, the most significant amount of memory can hence be saved by packing the two electron integrals $\tilde{g}_{pqrs}$ and additionally the four and five index amplitudes $t_{ijab}$ and $s^\alpha_{ijab}$. 
Overall, the memory requirements are then similar as compared to standard CC implementations.

\subsection{Implementation}
\label{subsec:val_of_imp}
The evaluation of integrals and the self-consistent-field iterations are carried out in the CFOUR program package.\cite{cfour,matthews2020coupled}
The QED-CC and EOM-QED-CC steps are carried out in the Qcumbre program prackage.\cite{qcumbre,hampe2017equation}
The algorithm exploits the block structure of matrices in all stages, i.e., the integral calculation, the QED self-consistent-field calculation, the integral transformation, and in the CC and EOM-CC algorithms.
The implemented algorithm was verified for single-polarization cavities by comparing results obtained using the \textit{$e^\text{T}$} program package\cite{eTpackage} for different systems, frequencies, and molecular orientations with respect to the polarization vector.
The algorithm was designed such that the single polarization appears as a special case of a cavity including an arbitrary number of EM modes.

\section{Results and Discussion}
\label{sec:calculations}
All calculations presented in this work were performed using a developer's version of the CFOUR\cite{cfour,matthews2020coupled} and Qcumbre\cite{qcumbre,hampe2017equation} program packages for the HF and CC steps, respectively.
All calculations employ point-group symmetry in all stages and were performed using a cc-pVTZ basis set\cite{dunning1989gaussian} unless stated otherwise.
The convergence criterion for all iterative steps was $10^{-8} E_\mathrm{h}$ or smaller. 
The molecular geometries have been optimized in absence of the cavity using the ORCA\cite{orca} program package on the DFT-B3LYP/def2-SVP level of theory.
The coupling strength was fixed  to $\lambda=0.05\;\text{a.u.}$ if not stated otherwise.
For all calculations the light-matter interaction is described with a single cavity frequency $\omega$ and interactions to other frequencies are neglected.

The algorithm exploits the block structure of matrices in all stages, i.e., the integral calculation, the QED self-consistent-field calculation, the integral transformation, and in the CC and EOM-CC algorithms.

The presented one-electron densities were calculated on a grid with at least $10^2$ grid points in each dimension and subsequently integrated over one coordinate. E.g., the integration of the $z$ coordinate yields
\begin{equation}
  \rho(x,y) = \int_{-\infty}^{\infty} \text{d}z \; \rho(x,y,z) 
  \label{eq:density-2d}
\end{equation}
which can be visualized in a contour matrix. 
An indicator to quantify the influence of the cavity on the molecule is obtained by integrating the change in the absolute density difference
\begin{equation}
  \Delta\rho = \int _\infty^\infty d^3r \mid \rho_\text{cav}(\boldsymbol{r}) - \rho_\text{ref}(\boldsymbol{r})\mid \;.
  \label{eq:den_diff}
\end{equation}
Here, $\rho_\text{cav}$ and  $\rho_\text{ref}$  are the one-electron densities within and without the cavity, respectively.  
This quantity can be understood as the fraction of an electron that is shifted in space and is given in the upper left corner of all density plots in the following. 

\subsection{Benzene}
\label{sec:aromatics}
Aromatic species are particularly interesting for polaritonic chemistry as they have isolated energetic low-lying excited states which can be easily addressed by tuning the cavity frequency.\cite{nobakht2024cavity}

We compare results for benzene in a linearly polarized (see also Ref. \citenum{barlini2024theory}) and unpolarized cavity, respectively. 
In principle, for calculations on benzene, the $D_{6 h}$ symmetry can be utilized by which time and memory savings are particularly advantageous.\cite{greiner2024expoiting}
However, in the presence of the cavity, the symmetry is reduced depending on the orientation of the polarization vectors with respect to the molecule.
Only in the cases where the polarization vector $\boldsymbol{\epsilon}$ or wave vector $\boldsymbol{k}$ is aligned perpendicularly to the molecular plane for the linear polarization and unpolarized cavity, respectively, the $D_{6 h}$ symmetry is maintained.
For other orientations, the symmetry is reduced, e.g., if the wave vector is aligned within the molecular plane, the symmetry is reduced to $D_{2 h}$. 
As most program packages, our algorithm is designed to only utilize real Abelian groups up to $D_{2h}$, as complex degeneracies require a more comprehensive algorithm.
\begin{figure}[t]
   \centering
   \includegraphics[width=1.\linewidth,trim=4 4 4 4,clip]{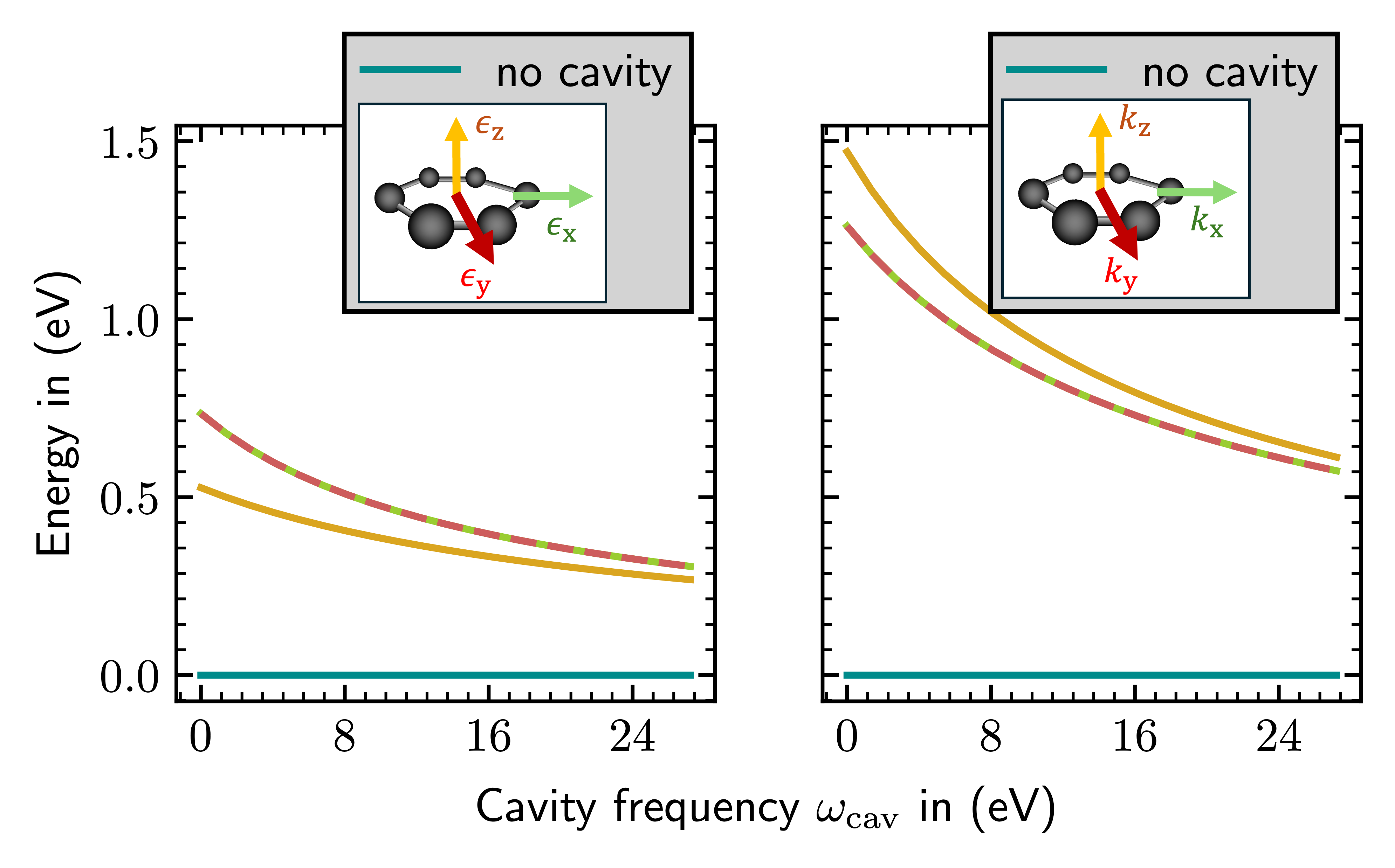}
   \caption{QED-CC ground-state energy difference of the benzene monomer with respect to the cavity frequency.
     Left: various orientations of benzene in a linearly polarized cavity; Right: various orientations of benzene in an unpolarized cavity.
     A coupling strength of $\lambda=0.1$ was employed.}
   \label{fig:benzen_energy}
\end{figure}

Figure (\ref{fig:benzen_energy}) shows how the energy of the benzene molecule changes with respect to the cavity frequency for a linearly polarized cavity (left) and an unpolarized cavity (right).
The preferred orientation of the molecule is obtained if the polarization vector $\boldsymbol{\epsilon}$ is aligned perpendicular to the molecular plane ($\epsilon_z$) or if the wave vector $\boldsymbol{k}$ is aligned within the molecular plane ($k_x$ and $k_y$), implying that one of the polarization vectors is aligned perpendicular to the molecular plane as well. 
This implies that the linearly polarized cavity stabilizes a single orientation ($\epsilon_z$) while the unpolarized cavity stabilizes the molecule oriented perpendicular to $k_z$ (which includes an infinite number of degenerate orientations with respect to a rotation around $\boldsymbol{k}$) as would be noticeable in a rotational spectrum.

The energy gap for the various orientations is the largest for the cavity frequency tending towards zero $\omega \rightarrow 0$ (for both, the linear and unpolarized cavity).
As the only term which explicitly depends on the cavity frequency is the bilinear coupling operator, the energy gap must be caused by the dipole self-energy as the gap is not vanishing when $\omega \rightarrow 0$.
This limit should however not be confused with the cavity-free case, as the coupling strength remains constant.
Hence, this limit must be viewed with caution, as in a physical system with an arbitrarily large cavity also the coupling strength $\lambda$  will go to zero for $\omega \rightarrow 0$.
Increasing the cavity frequency results in a $ E \propto - \sqrt{\omega}$ dependency, which stems from the prefactor of the bilinear coupling term. 
To influence ground-state properties with an optical cavity, it is therefore not as important to tune the cavity in resonance with excited states as it is for the formation of upper and lower polaritonic structures.
\begin{figure}[t]
   \centering
   \includegraphics[width=1.\linewidth,trim=4 4 4 4,clip]{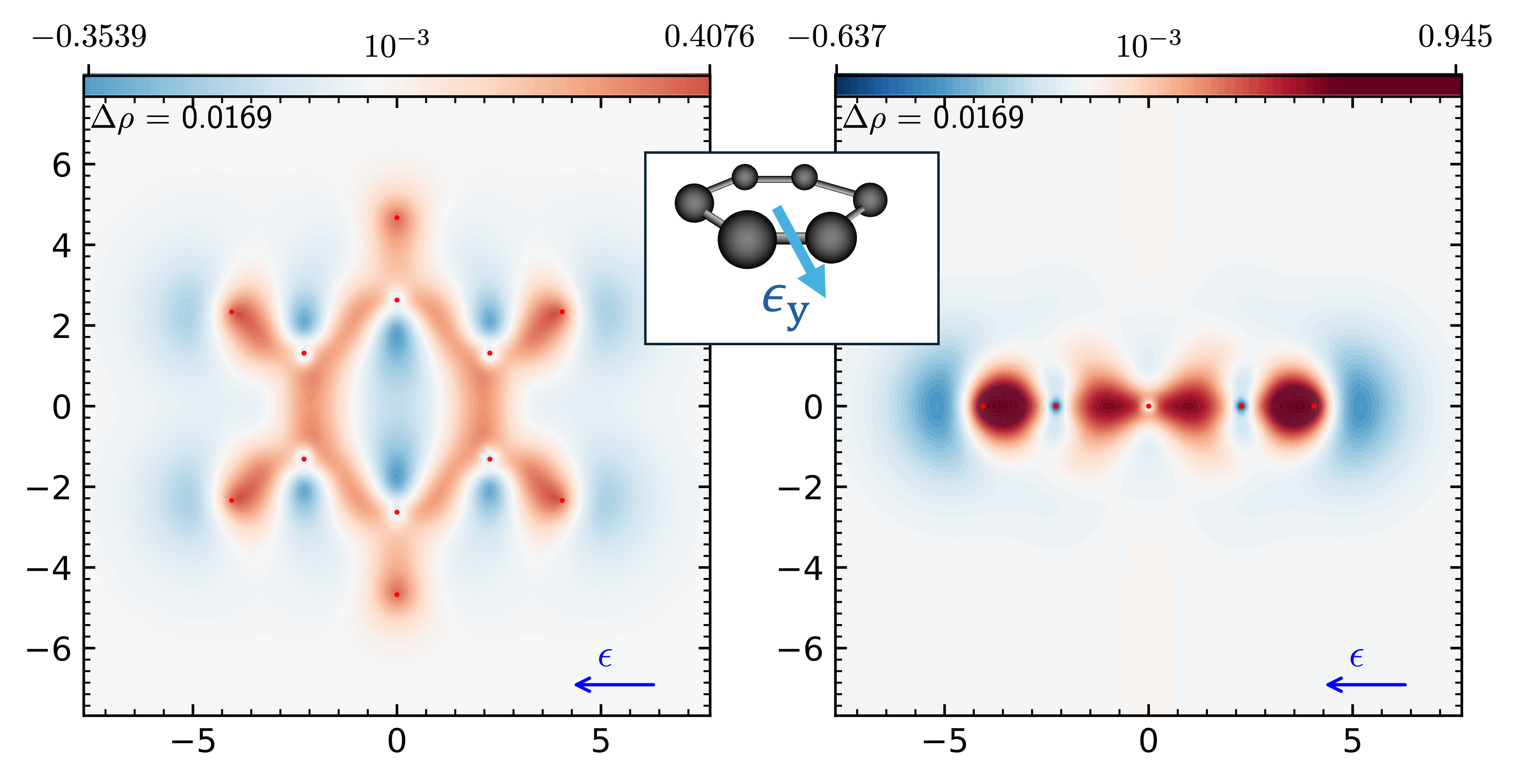}
   \includegraphics[width=1.\linewidth,trim=4 4 4 4,clip]{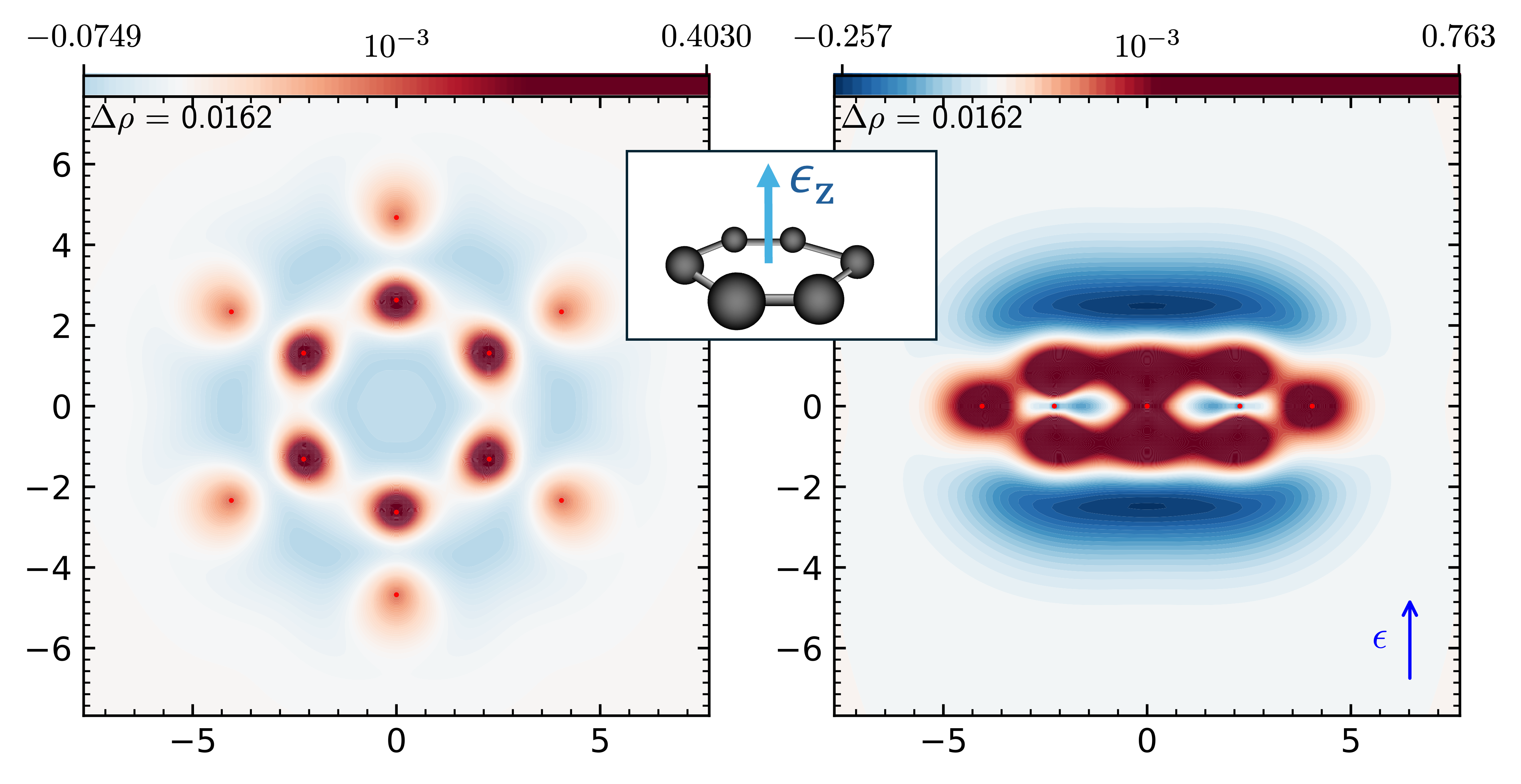}
   \caption{Correlated one-electron density differences for benzene in a linearly polarized cavity. 
   Left and right plots are different perspectives on the same density difference.
   The upper plots show the $\boldsymbol{\epsilon}$ vector aligned in the molecular plane and the lower plots show the $\boldsymbol{\epsilon}$-vector normal to the molecular plane.}
   \label{fig:benzen_difference_lin}
\end{figure}

\begin{figure}[t]
   \centering
   \includegraphics[width=1.\linewidth,trim=4 4 4 4,clip]{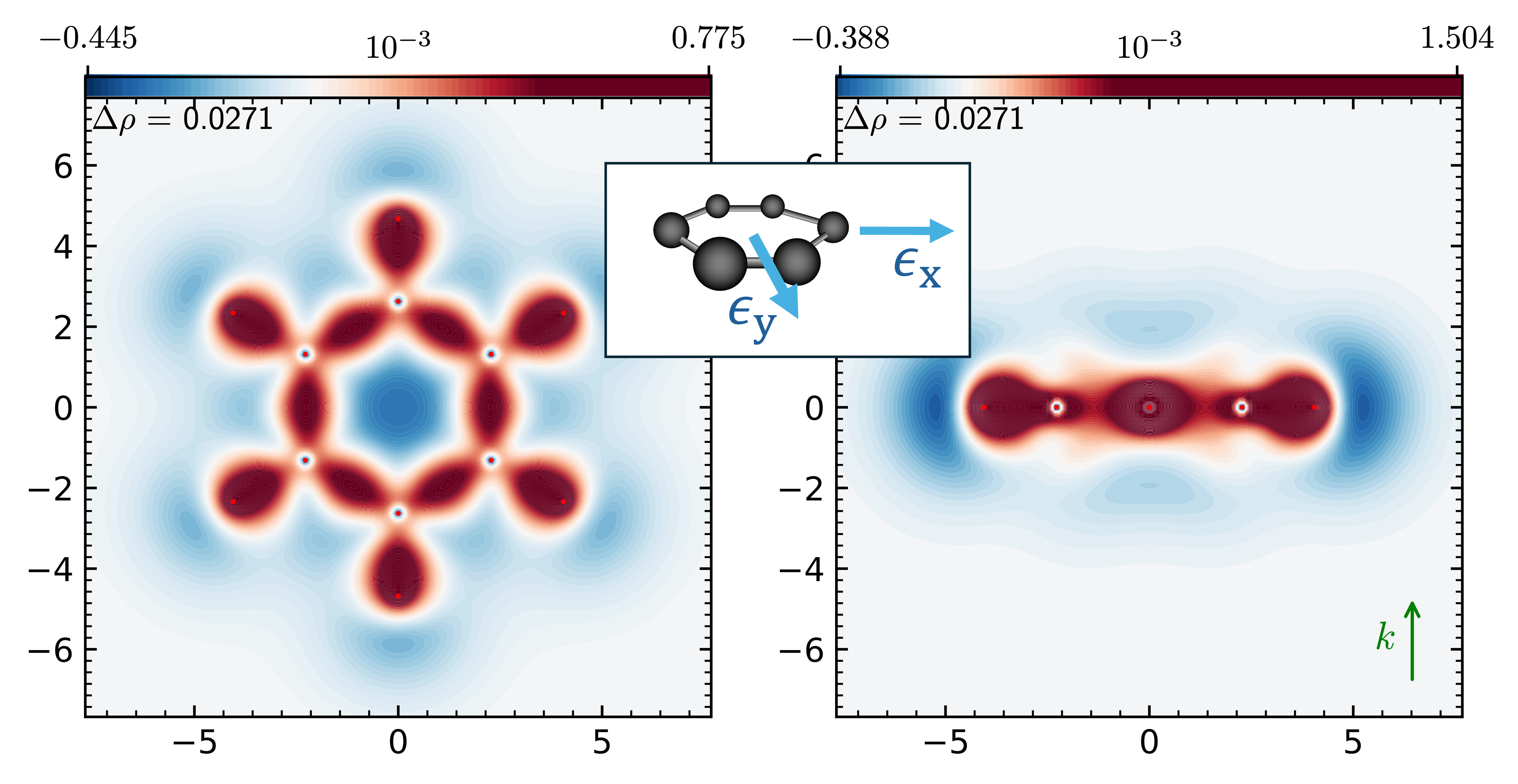}
   \includegraphics[width=1.\linewidth,trim=4 4 4 4,clip]{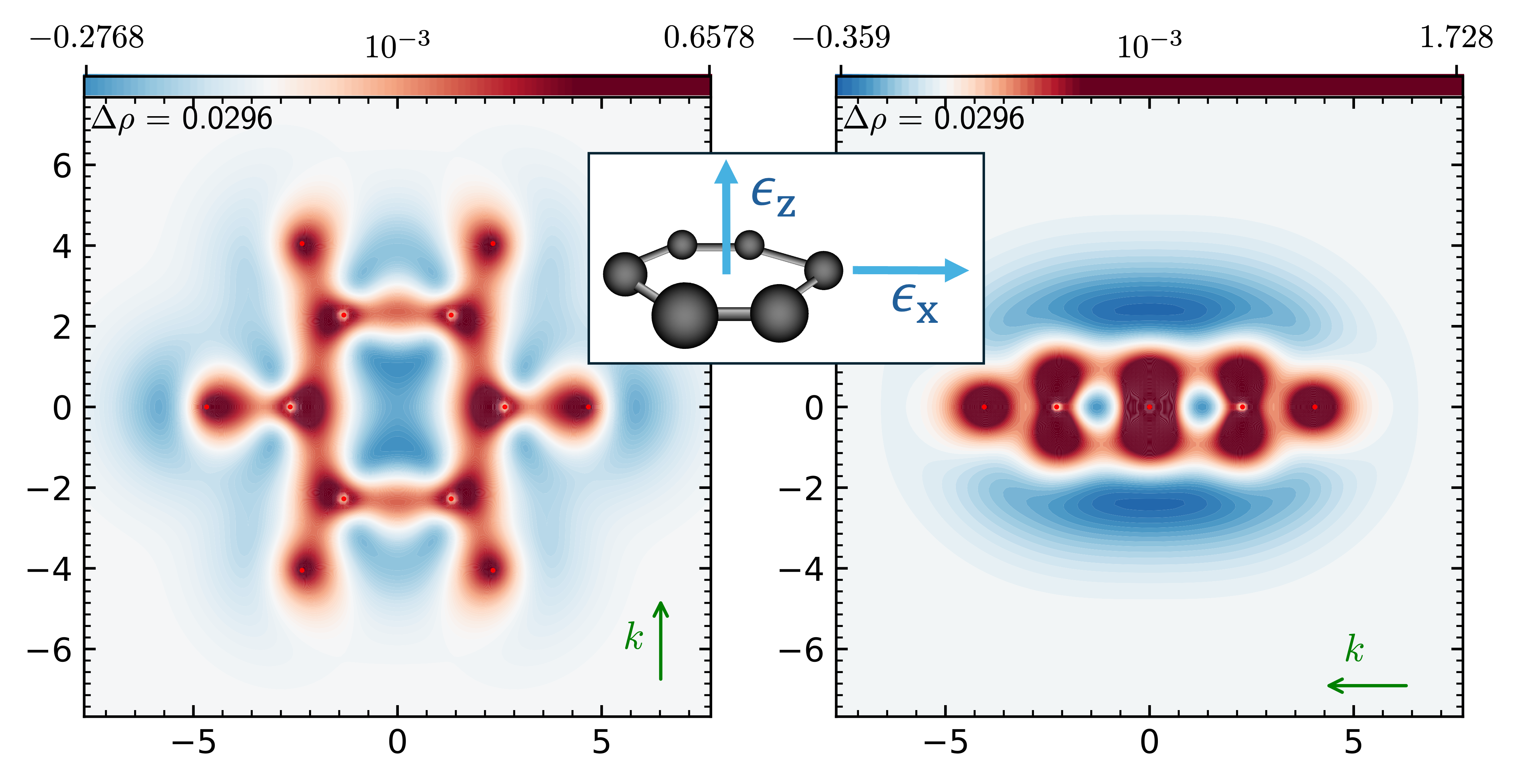}
   \caption{Correlated one-electron density difference for a benzene molecule in an unpolarized cavity. 
   Left and right plots are different perspectives on the same density difference.
   Lower plots show the $\boldsymbol{k}$ vector aligned in the molecular plane and the upper  plots show the $\boldsymbol{k}$-vector normal to the molecular plane.}
   \label{fig:benzen_difference}
\end{figure}

The reduction in symmetry to $D_{2h}$ is visible in the one-electron density differences in Fig (\ref{fig:benzen_difference}, top left) for the more stable orientation in which the $\boldsymbol k$ vector lies in the molecular plane.
Note that the left and right plots show different perspectives for the same orientation of the molecule within the cavity. 
In the plots, a buildup of electron density with respect to the cavity-free case is indicated in red while a decrease is shown in blue.  
In the cavity, the density becomes more localized in the C-C bonds and on the hydrogen atoms, see also Ref. \citenum{flick_2018}, where density differences have been reported at the QEDFT level of theory. 
We note that, as expected, the density is mostly shifted along the polarization vectors. 

In the case where the $\boldsymbol{k}$-vector is aligned in the molecular plane, one polarization vector is oriented normal to the molecular plane and the density is partly shifted into the $\pi$-system (Fig. \ref{fig:benzen_difference} top right), while for the case where both polarization vectors are localized in the molecular plane, the density is shifted into the $\sigma$-system (Fig. \ref{fig:benzen_difference} bottom).
Mostly, for the ground-state density, the difference $\Delta \rho$ is very small. For benzene, it ranges between $10^{-2}$ and $10^{-1}$, showing that the influence of an optical cavity on the ground-state is relatively small.

Comparing the benzene molecule in the linearly and unpolarized cavities, respectively, reveals some differences (see Figs. \ref{fig:benzen_difference_lin} and \ref{fig:benzen_difference}).
First of all, the overall trends of the linearly polarized cavity agree with the results presented in Ref. \citenum{barlini2024theory}. In particular, it is found that electron density is accumulated in the $\pi$-system for a polarization vector perpendicular to the molecule. 

When comparing to the results for unpolarized cavities, one has to keep in mind that the resulting point groups may differ: 
For linearly polarized cavities, the $D_{6h}$ symmetry is preserved for a polarization vector $\boldsymbol\epsilon$ oriented perpendicular to the molecular plane. 
Aligning the polarization vector within the molecular plane reduces the symmetry to $D_{2h}$. 
This is contrary to the unpolarized cavity, where the $D_{6h}$ orientation is preserved for the polarization vectors lying in the molecular plane (and $\boldsymbol k$ perpendicular to it). 
Thus, even if the systems have the same point-group symmetry, they describe different physical situations.
Furthermore, comparing the density differences $\Delta \rho$ in Fig \ref{fig:benzen_difference} and \ref{fig:benzen_difference_lin}, reveals that the change in the density is for an unpolarized cavity about twice as high as for a linearly polarized cavity.
This is mainly due to the fact, that the molecule is coupled by two perpendicularly polarized modes to the electric field, so the coupling effects are twice as strong. 

\subsection{Fluorobenzene}
\label{subsec:fluorobenzene}
\begin{figure}[t]
   \centering
   \includegraphics[width=1.\linewidth,trim=4 4 4 4,clip]{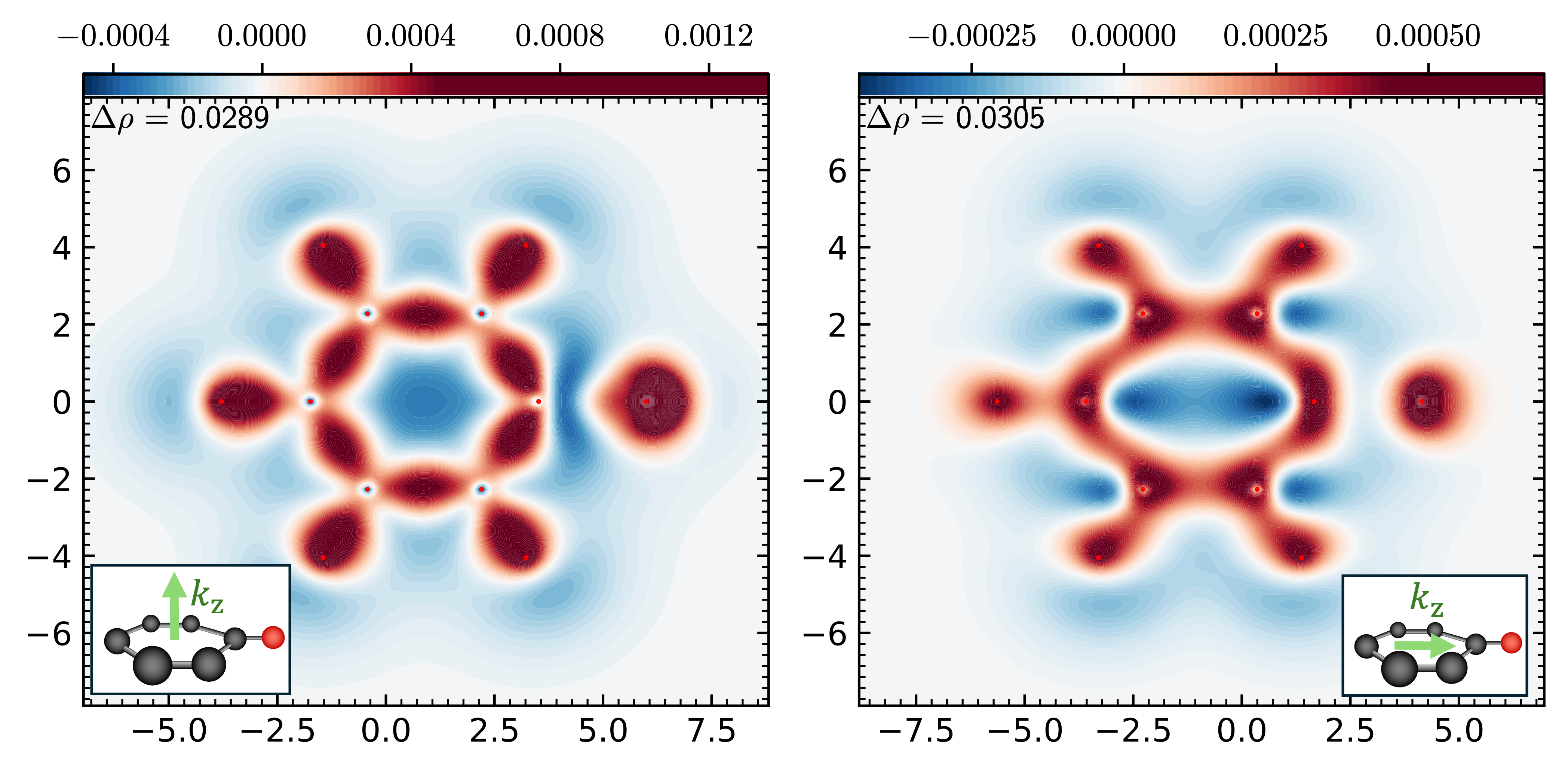}
   \caption{Correlated one-electron density differences for fluorobenzene in two orientations within an unpolarized cavity. 
            }
   \label{fig:fluor_benz_difference}
\end{figure}

For the fluorobenzene molecule, qualitative differences as compared to benzene are found: 
Fig. \ref{fig:fluor_benz_difference} shows that the cavity can also be responsible for removing electrons from a bond, here the F-C bond. 
Hence, both accumulation and depletion of electron density in bonding regions may occur. 
The density shifts are mainly perpendicular to the $\boldsymbol k$ vector along the polarization vectors as can be seen by comparing the case where the $\boldsymbol k$ vector is perpendicular to the molecular plane and where the $\boldsymbol{k}$ vector lies within the molecular plane.
If the $\boldsymbol{k}$-vector is aligned within the molecular plane and parallel to the F-C bond (hence the polarization vectors are perpendicular to the bond) the electron density in the bond is almost unaffected.
When the $\boldsymbol{k}$-vector is perpendicular to the F-C bond (hence one polarization vector can be aligned parallel to it) the electron density in the bond is significantly reduced.
A similar behavior is encountered in a linearly polarized cavity when the polarization vector is oriented along the F-C bond. (see SI, Fig.\ref{fig:fluor_benz_difference_lin} )
\\
\subsection{Azulene}
\label{subsec:azulene}
\begin{figure}[t]
   \centering
   \includegraphics[width=1.\linewidth,trim=4 4 4 4,clip]{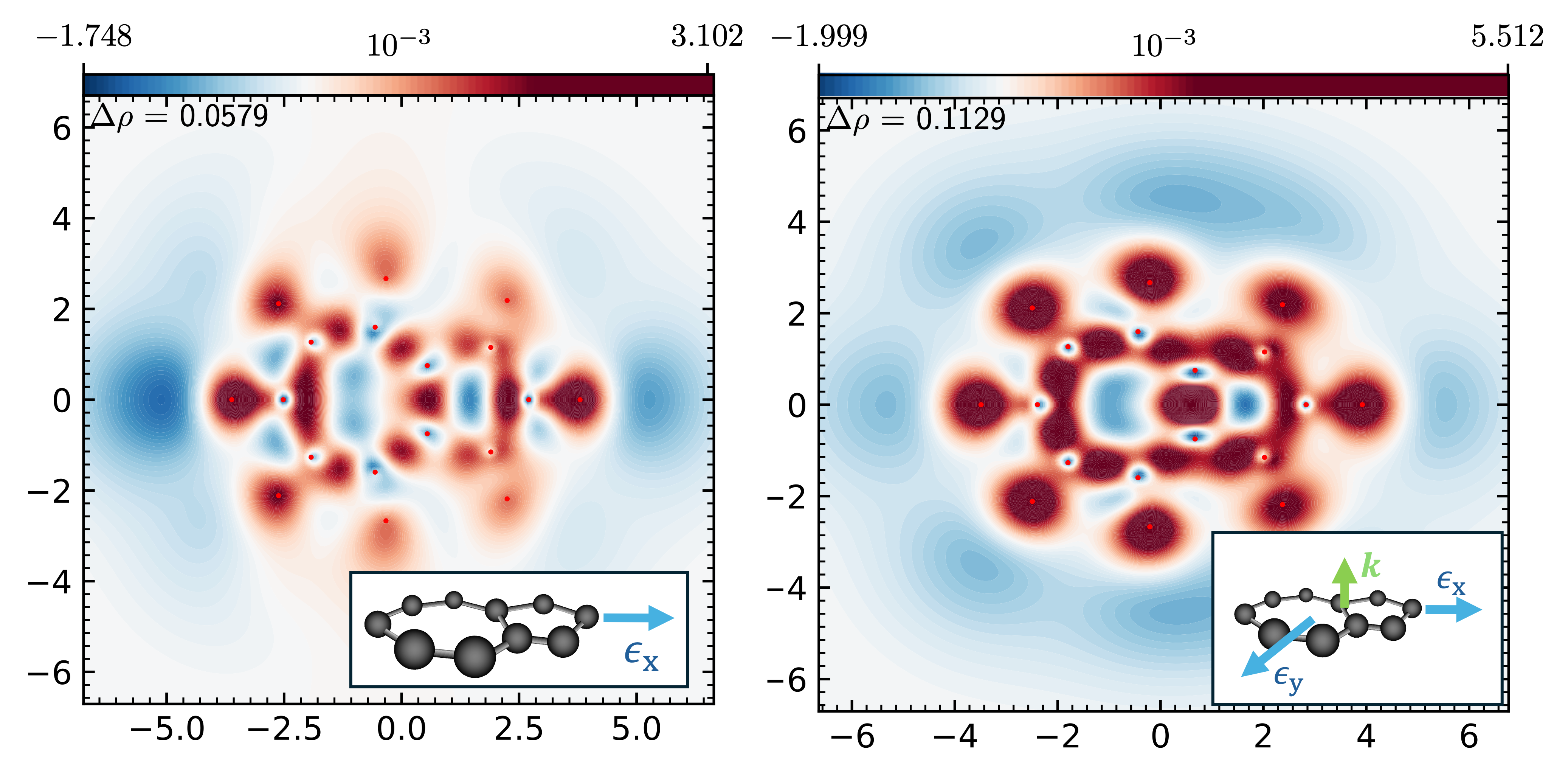}
   \caption{Correlated one-electron density differ-
ences for the azulene molecule in a linearly polarized (left) and unpolarized cavity (right), respectively. 
   The cavity frequency was set to $\omega=2.41\;eV$ and the coupling strength $\lambda=0.08 \text{ a.u}$.  }
   \label{fig:azulene}
\end{figure}

The azulene molecule 
has been investigated at the QEDFT level of theory for a linearly polarized cavity by Flick et al.\cite{flick_2018}
This allows for a comparison between QED-CC and QEDFT predictions. 
Fig (\ref{fig:azulene}) depicts the azulene molecule in a linearly (left) and unpolarized cavity (right) for a coupling strength of $\lambda=0.08\;\text{a.u.}$ and a frequency of $\omega=2.41 \;eV$. 
These parameters are consistent with those used in Ref. \citenum{flick_2018}, except for the molecular geometry, which may differ slightly.
 Flick et al. presented that the azulene molecule develops a rich fine structure in the one-electron density differences within a linearly polarized cavity. 
At a first glance, the density difference for the linearly polarized cavity appears similar to the results presented at the DFT level of theory, in particular for the results obtained for the Krieger-Li-Iafrate approximation in Ref. \citenum{flick_2018}. 
Most of the electron density is accumulated at the Hydrogen atoms oriented along the polarization vector to both ends of the azulene molecule and the $\sigma$ C-C bonds. 
We do not find an accumulation of electron density in the $\alpha$ position in the heptatrien ring as suggested by the DFT results with an optimized-effective potential (OEP), see also Ref. \citenum{flick_2018}.
We note that corresponding density difference plots obtained at the HF level of theory yield qualitatively similar results to those obtained at the QED-CC level (see SI, Figs. \citenum{flick_2018}). This suggests that the influence of the cavity on the electron density is mainly due to the dipole self-energy. 
  
As expected, in an unpolarized cavity, see Fig. \ref{fig:azulene} right, where the $\boldsymbol{k}$-vector of the cavity is aligned perpendicular to the molecule with the two polarization vectors lying in the molecular plane, the electron density is shifted in a symmetric manner and localized at the Hydrogen atoms and the C-C $\sigma$ bonds. 

\subsection{Symmetries of excited states within polaritonic EOM-CC theory} 
\label{subsec:excited_states}
In the uncoupled system ($\lambda_\alpha = 0$) the excitation operator $\hat{R}$ yields in the simplest case either an electronic excitation $\hat{R}_\mu\ket{0,0} = \ket{\mu,0}$ or a photonic creation  $\hat{R}_\nu \ket{0,0} = \ket{0,\nu}$.
In the following, the excited states are referred to by the irreducible representation they belong to. 
The electronic ground state corresponds to the Fermi vacuum with symmetry $\Gamma_G$, while the photonic ground state corresponds to the physical vacuum and is hence totally symmetric $\ket{0,0} = \ket{\Gamma_G,\Gamma_1}$.
This scheme can be adapted for electronic and photonic excited states. For the electronic states we have $\ket{\mu,0}=\ket{\Gamma_E,\Gamma_1}$.
For the photonic excited states two notations are used   addressing either the number of photons or  the irreducible representation of the photonic mode. I.e., an excitation with one photon will be denoted as $\ket{0,\nu} = \ket{\Gamma_G, 1}$ or as $\ket{0,\nu} = \ket{\Gamma_G, \Gamma_\alpha}$, respectively.
Increasing the coupling strength mediates interactions via the bilinear coupling operator and leads to the formation of an upper and lower polariton $\ket{P_+}$ and $\ket{P_-}$.
In the simplest case of just two interacting states, the upper and lower polaritons are given by a linear combination of the uncoupled states:
\begin{equation}
   \begin{split}
   \ket{P_+} & = c_1 \ket{\Gamma_E,\Gamma_1} + c_2 \ket{\Gamma_G,\Gamma_\alpha}\;,\\
   \ket{P_-} & = c_3 \ket{\Gamma_E,\Gamma_1} - c_4 \ket{\Gamma_G,\Gamma_\alpha}\;.
   \end{split}
   \label{eq:polariton}
\end{equation}
The  coupling occurs if the uncoupled electronic excited state $\ket{\Gamma_{E}, \Gamma_1}$ and the cavity excitation $\ket{\Gamma_{G},\Gamma_\alpha}$ belong to the same irreducible representation.
The states hence only couple if $ \Gamma_1 \in \Gamma_G \otimes \Gamma_\alpha \otimes \Gamma_E$. 
The coefficients $c_i$ are heavily influenced by the frequency and coupling strength of the cavity. The formation of the upper and lower polariton can be viewed as the result of an avoided crossing of an electronic excited state and an excitation of the EM field.\cite{koehn2007can,csehi2019quantum}
Of course, if more states of the same irreducible representation are in the  energetic vicinity of the cavity frequency, they may contribute to the interaction. 

\begin{figure}[t]
   \includegraphics[width=1.0\linewidth]{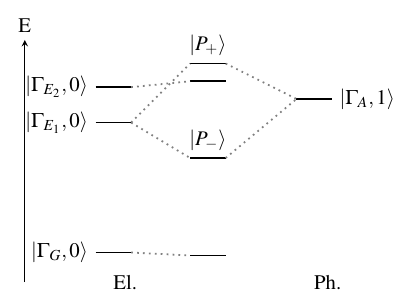}
\caption{Exemplary energy landscape for an electronic three level system in presence of a single photonic mode.
  The irreducible representations of the electronic ground state is $\Gamma_G$ and for the two excited states $\Gamma_{E_1}$ and $\Gamma_{E_2}$.
The irreducible representation of the photon is $\Gamma_\mathrm{ph}$.
For the depicted system $\Gamma_G \otimes \Gamma_\mathrm{ph} = \Gamma_{E_1} \ne \Gamma_{E_2}$ }
\label{fig:polariton1}
\end{figure}

Note that even though the notation  $\ket{\Gamma_G,0}$ seems to imply that no photons are present, the exponential of the cluster operator $\text{e}^{\hat{Q}}$ is present in the similarity-transformed Hamiltonian. Hence, the excited state within polaritonic CC theory is given as  
\begin{equation}
    \ket{\Psi_\text{exc}} = \text{e}^{\hat{Q}}\hat{R}\ket{\Gamma_G,0} = \text{e}^{\hat{Q}} \ket{\Gamma_E,\Gamma_\alpha} .
\end{equation}
\\
An exemplary energy diagram for the formation of polaritons in a three level system of electronic ground state $G$, and excited states $E_1$ and $E_2$ is shown in Fig. \ref{fig:polariton1}.
Note that for each electronic state $\ket{\Gamma_E,0}$, there exists one photonic state $\ket{\Gamma_E,1}$ shifted in energy by the cavity frequency $\omega_\alpha$.
Only the photonic state $\ket{\Gamma_G,1}$ is depicted, while in theory also the photonic states $\ket{\Gamma_{E_1},1}$ and $\ket{\Gamma_{E_2},1}$ exist, but are assumed to be  energetically too high for any significant coupling. 
\\
An intricate feature in the coupling emerges 
when the photonic states are excited with more than one photon. 
The Hamiltonian in Eq. (\ref{eq:ham}) only couples between states which differ effectively by not more than one photon.
This means that in a basis of Slater-determinants, there is no coupling between an electronically singly excited determinant $\ket{\Gamma_E,0}$  and the doubly excited photonic excited ground state $\ket{\Gamma_G,2}$.
Therefore,
\begin{equation}
   \begin{split}
     0 = \braket{\Gamma_G,2|\hat{H}|\Gamma_E,0} \;.
   \end{split}
   \label{eq:coupling}
\end{equation}
The similarity-transformed Hamiltonian $\hat{\tilde{H}}$, on the other hand, contains multiple photonic excitations and hence also states that differ by more than one photon are in principle coupled
\begin{equation}
   \begin{split}
   0 \ne \braket{\Gamma_G,2|\hat{\tilde{H}}|\Gamma_E,0} 
   \end{split}
   \label{eq:transition_similarity}.
\end{equation}
These couplings are, however, expected to be very weak, as they are a result of the correlation contributions that involve photonic excitations. 
Also, different truncation schemes can lead to different light-matter couplings.
E.g. in the CCSD-12-SD truncation, the states $\ket{\Gamma_E,0}$ and $\ket{\Gamma_G,2}$ are not directly coupled, i.e., 
\begin{equation}
   \begin{split}
     0 = \braket{\Gamma_G,2|\hat{\tilde{H}}_\text{CCSD-12}|\Gamma_E,0} \;.
   \end{split}
\end{equation}
This is due to the fact that $\hat{\tilde{H}}_\text{CCSD-12}$ has in the given truncation no term with two-photon creations and a single electronic deexcitation which would require at least the $\hat \Gamma_3$ or the $\hat{S}_{1}^{2}$ operator in the cluster operator $\hat{Q}$:
\begin{equation}
   \begin{split}
   \hat{\Gamma}_3 &= \frac{1}{3!} \sum_{\alpha \beta \lambda} \gamma^{\alpha\beta\lambda} \hat{\alpha}^\dagger \hat{\beta}^\dagger \hat{\lambda}^\dagger\\
   \hat{S}_1^2 &= \frac{1}{2!} \sum_{ai \alpha \beta} s_{ai}^{\alpha\beta} \hat{\alpha}^\dagger \hat{\beta}^\dagger \hat{a}^\dagger\hat{i}\\
   \end{split}
   \label{eq:transition_similarity}
\end{equation}
Both operators can lead to direct couplings of the states $\ket{\Gamma_E,0}$ and $\ket{\Gamma_G,2}$ which can be seen from the commutators
\begin{equation}
   \begin{split}
  &  \braket{\Gamma_G,2| [ \hat{H},\hat{\Gamma}_3 ] |\Gamma_E,0} \leftarrow 
  \begin{minipage}{0.2\linewidth}\includegraphics[width=1.0\linewidth]{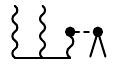}\end{minipage} = \\
  & = \sum_{a i \alpha \beta} \sum_{\lambda} \tilde{d}_{ai}^\lambda \gamma^{\alpha\beta\lambda} \braket{\Gamma_G,2| \hat{i}^\dagger \hat{a}\hat{\alpha}^\dagger \hat{\beta}^\dagger |\Gamma_E,0} 
   \end{split}
\end{equation}
and
\begin{equation}
   \begin{split}
  &  \braket{\Gamma_G,2| [ \hat{H},\hat{S}_1^2 ] |\Gamma_E,0} \leftarrow 
  \begin{minipage}{0.2\linewidth}\includegraphics[width=1.0\linewidth]{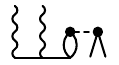}\end{minipage} = \\
  & = \sum_{a i \alpha \beta} \sum_{em} \tilde{g}_{miea} s_{me}^{\alpha\beta} \braket{\Gamma_G,2| \hat{i}^\dagger \hat{a}\hat{\alpha}^\dagger \hat{\beta}^\dagger |\Gamma_E,0} \;.
   \end{split}
\end{equation}
For unpolarized cavities, a similar argumentation holds for states of the type $\ket{\Gamma_E,0}$ and $\ket{\Gamma_G, \Gamma_\alpha \Gamma_{\bar{\alpha}}}$ where two photons are localized in perpendicularly polarized modes, respectively.
The same is expected for the states  $\ket{\Gamma_{E_1}, \Gamma_\alpha}$ and $\ket{\Gamma_{E_2}, \Gamma_{\bar{\alpha}}}$, even if they are both excited with one photon. 
But, as the photon polarization is orthogonal, the states effectively differ in two photons. 
\\
This highlights the challenge of selecting an appropriate cluster operator, as the presence or absence of avoided crossings can depend on the chosen truncation scheme.
In section (\ref{subsec:H2}) these issues will be discussed for the example of H$_2$ showing  the importance of these considerations already for a small test system. 

\subsection{H$_2$ in linear and   unpolarized cavities}
\label{subsec:H2}
\begin{figure}[t]
   \centering
   \begin{minipage}{1.0\linewidth}
     \includegraphics[width=\textwidth,trim=4 4 4 4,clip]{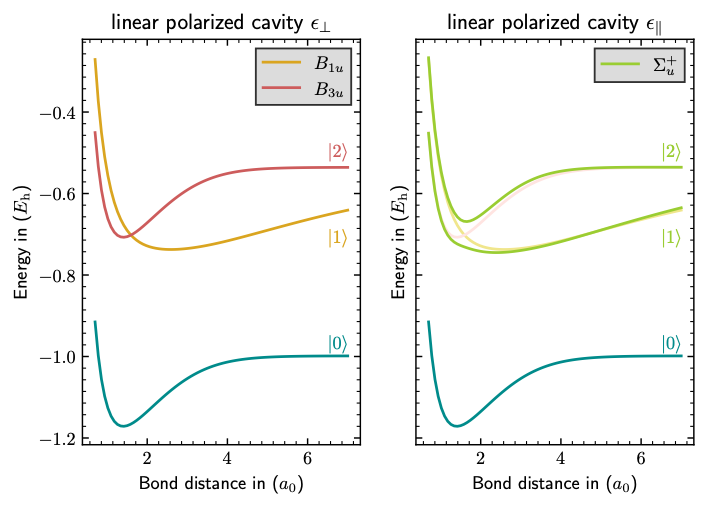}
   \end{minipage}
   \begin{minipage}{1.0\linewidth}
    \centering
    \begin{tabular}{c c c c}
       \hline
       \hline
       state & \multicolumn{1}{|c}{sym} & \multicolumn{2}{|c}{coupling}\\
       \hline
       \multicolumn{4}{c}{parallel}   \\
       \hline
       $\ket{0}$ & $\Sigma_g^+ $ & $ \ket{\Sigma_g^+,0}  $&$ \ket{\Sigma_g^+, \Sigma_g^+} $    \\
       $\ket{1}$ & $\Sigma_u^+ $ & $ \ket{\Sigma_u^+,0}  $&$ \ket{\Sigma_u^+,\Sigma_g^+}$ \\ 
       $\ket{2}$ & $\Sigma_u^+ $ & $ \ket{\Sigma_g^+, 1} $&$ \ket{\Sigma_g^+,\Sigma_u^+}$ \\
        \hline

       \multicolumn{4}{c}{perpendicular} \\
       \hline
       $\ket{0}$ &  $A_{g}  $ &  $\ket{A_{g\;},0} $&$ \ket{A_g, A_g} $ \\
       $\ket{1}$ &  $B_{1u} $ &  $\ket{B_{1u},0}  $&$ \ket{B_{1u},A_g}$ \\ 
       $\ket{2}$ &  $B_{3u} $ &  $\ket{A_g,1}     $&$ \ket{A_g, B_{3u}} $\\
        \hline
        \hline
    \end{tabular}
    \label{tab:h2_un_lin}
   \end{minipage}
   \caption{Low-lying singlet states of H$_2$ in a linearly polarized cavity with the polarization vector aligned parallel (green) and perpendicular (red) with respect to the molecular axis.
            Only selected states of \textit{ungerade} parity are shown. 
            The cavity frequency was set to $12.48\; eV$ and the coupling strength to $0.05 \text{ a.u.}$
            The table shows different notations for the polaritonic states in the parallel as well as perpendicular orientations.
            }
   \label{fig:h2_un_lin}
\end{figure}

In the following we study the effects of a linearly polarized and an unpolarized cavity on the H$_2$ molecule.
We note that the polaritonic ground state does not change significantly in the cavity\cite{haugland2020coupled, haugland2021intermolecular} and hence we only discuss the polaritonic excited states. 
Low-lying singlet states of H$_2$ in a linearly polarized cavity are depicted in Fig.  \ref{fig:h2_un_lin}, where the ground state and the two lowest excited states of \textit{ungerade} parity are shown.
The molecule is aligned parallel to the $z$-axis while the cavity orientation changes. 
The different notations for the total irreducible representation of the states as well as the contributions from the electronic and photonic parts are listed in the table within Fig. \ref{fig:h2_un_lin}.
Fig.  \ref{fig:h2_un_lin} (left) shows that no splitting into upper and lower polariton occurs for the perpendicular orientation of the polarization vector as the electronically excited state $\ket{1}$ and the photonic excited state $\ket{2}$ are characterized by different irreducible representations.
Note that the $\ket{2}$ state is the ground-state shifted by the frequency of the cavity $\omega_\alpha$ which has a strong photonic character.
For the parallel orientation of the polarization vector (right), however, the states $\ket{1}$ and $\ket{2}$ belong to the same irreducible representation $\Sigma_u^+$ and a coupling resulting in an upper and lower polariton is observed.
\begin{figure}[t!]
   \centering
   \begin{minipage}{1.0\linewidth}
     \includegraphics[width=\textwidth,trim=4 4 4 4,clip]{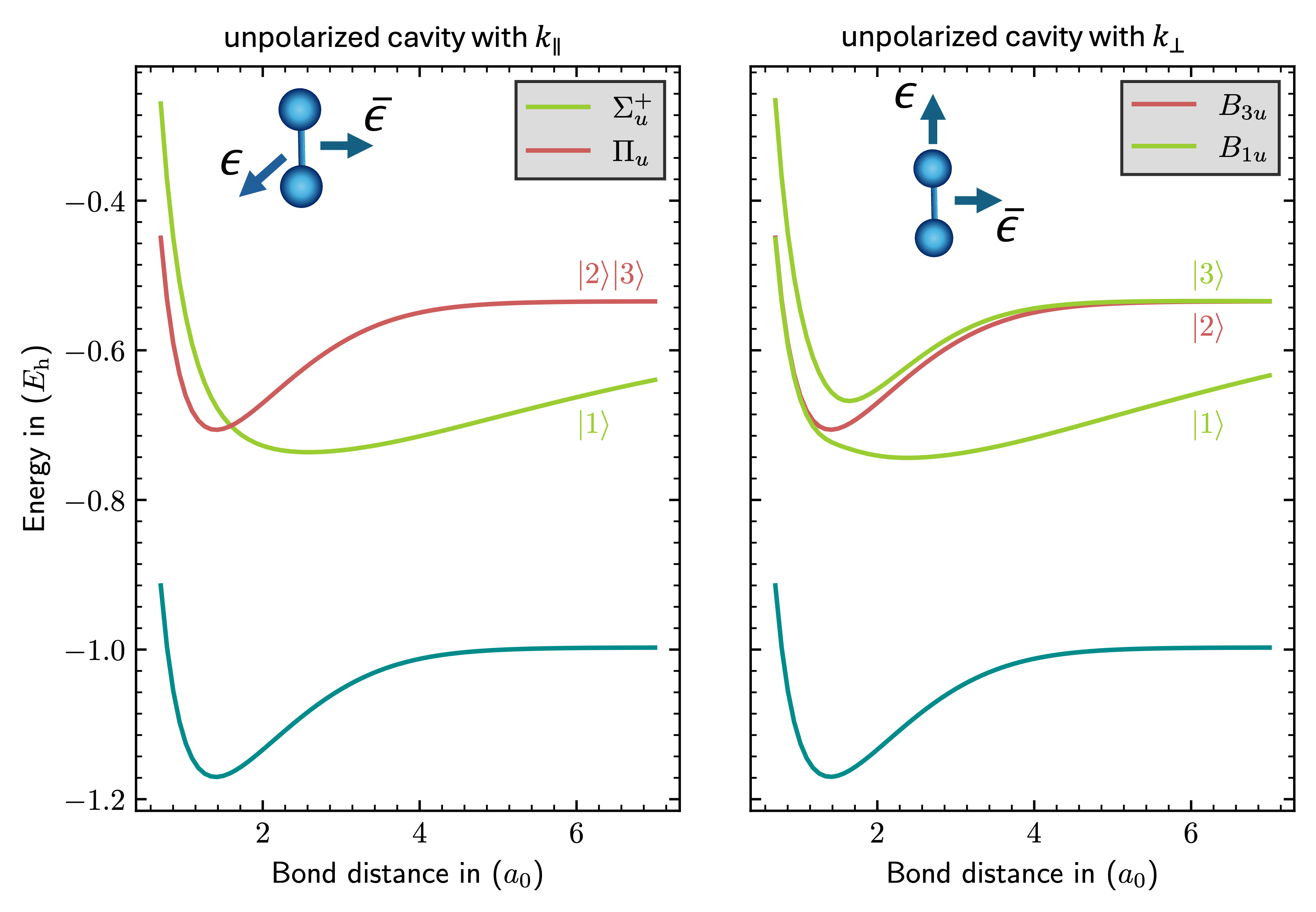}
   \end{minipage}
   \begin{minipage}{1.0\linewidth}
    \centering
    \begin{tabular}{c c c c}
       \hline
       \hline
       state & \multicolumn{1}{|c}{sym} & \multicolumn{2}{|c}{coupling}\\
       \hline
       \multicolumn{4}{c}{parallel}   \\
       \hline
       $\ket{1}$ & $\Sigma_u^+ $ & $ \ket{\Sigma_u^+,0}  $&$ \ket{\Sigma_u^+,\Sigma_g^+}$ \\ 
       \multirow{2}{*}{$\ket{2} \ket{3}$}  & \multirow{2}{*}{$\Pi_u $}  & $ \ket{\Sigma_g^+, 1} $&$ \ket{\Sigma_g^+,\Pi_u^-}$ \\
       &  & $ \ket{\Sigma_g^+, \bar{1}} $&$ \ket{\Sigma_g^+,\Pi_u^+}$ \\
        \hline

       \multicolumn{4}{c}{perpendicular} \\
       \hline
       $\ket{1}$ &  $B_{1u} $ &  $\ket{A_{g},1}  $&$ \ket{A_{g},B_{1u}}$ \\ 
       $\ket{2}$ &  $B_{3u} $ &  $\ket{A_{g},\bar{1}}  $&$ \ket{A_{g},B_{3u}}$ \\ 
       $\ket{3}$ &  $B_{1u} $ &  $\ket{B_{1u},0}  $&$ \ket{B_{1u}, A_g}$ \\ 
        \hline
        \hline
    \end{tabular}
   \end{minipage}
   \caption{Low-lying singlet states of H$_2$ in a symmetrically polarized cavity with the wave-vector aligned parallel (left) and perpendicular (right) with respect to the molecular axis.
            Only selected states of \textit{ungerade} parity are shown to focus on the formation of the upper and lower polariton.
            The cavity frequency was set to $12.48 eV$ and the coupling strength to $0.05 \text{ a.u.}$
            Below: different notations for the polaritonic states of $^1$H$_2$ in parallel as well as perpendicular orientation.
            }
   \label{fig:h2_un_sym}
\end{figure}

\begin{figure}[t!]
   \centering
   \begin{minipage}{1.\linewidth}
     \includegraphics[width=\textwidth,trim=4 4 4 4,clip]{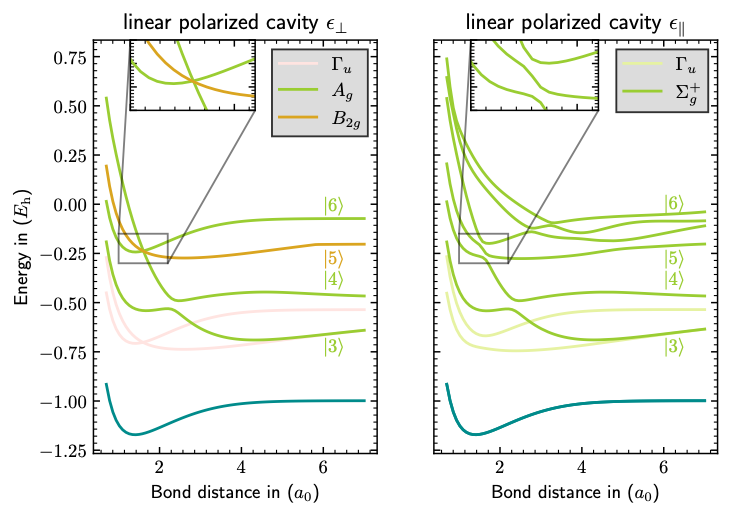}
    \end{minipage}
    \centering
   \begin{minipage}{1.\linewidth}
    \centering
    \begin{tabular}{c c c c}
       \hline
       \hline
       state & \multicolumn{1}{|c}{sym} & \multicolumn{2}{|c}{coupling}\\
       \hline
       \multicolumn{4}{c}{perpendicular}   \\
       \hline
       $\ket{3}$ &  $A_{g}  $ &  $\ket{A_{g\;},0} $&$ \ket{A_g, A_g} $ \\
       $\ket{4}$ &  $A_{g}  $ &  $\ket{A_{g\;},0} $&$ \ket{A_g, A_g} $ \\
       $\ket{5}$ &  $B_{2g}  $ &  $\ket{B_{1u},1} $&$ \ket{B_{1u}, B_{3u}} $ \\
       $\ket{6}$ &  $A_{g} $ &  $\ket{A_g,2}     $&$ \ket{A_g, A_g} $\\
        \hline

       \multicolumn{4}{c}{parallel} \\
       \hline
       $\ket{3}$ & $\Sigma_g^+ $ & $ \ket{\Sigma_g^+,0}  $&$ \ket{\Sigma_g^+, \Sigma_g^+} $    \\
       $\ket{4}$ & $\Sigma_g^+ $ & $ \ket{\Sigma_g^+,0}  $&$ \ket{\Sigma_g^+, \Sigma_g^+}$ \\ 
       $\ket{5}$ & $\Sigma_g^+ $ & $ \ket{\Sigma_u^+, 1} $&$ \ket{\Sigma_u^+,\Sigma_u^+}$ \\
       $\ket{6}$ & $\Sigma_g^+ $ & $ \ket{\Sigma_g^+, 2} $&$ \ket{\Sigma_g^+,\Sigma_u^+}$ \\
        \hline
        \hline
    \end{tabular}
    \end{minipage}
   \caption{Low-lying singlet states of H$_2$ in a linear polarized cavity with the polarization vector aligned perpendicular (left) and parallel (right) with respect to the molecular axis.
            Only selected states of \textit{gerade} parity are shown to focus on the formuation of the upper and lower polariton.
            In faded colors also states of \textit{ungerade} parity are shown for comparison.
            The cavity frequency was set to $12.48 eV$ and the coupling strength to $0.05      \text{ a.u.}$
            Below: different notations for the polaritonic states of $^1$H$_2$ in parallel as well as perpendicular orientation.
            }
   \label{fig:h2_ger_lin}
\end{figure}
\begin{figure*}[h!]
    \centering
  \begin{minipage}{0.89\textwidth}
     \includegraphics[width=\textwidth,trim=4 4 4 4,clip]{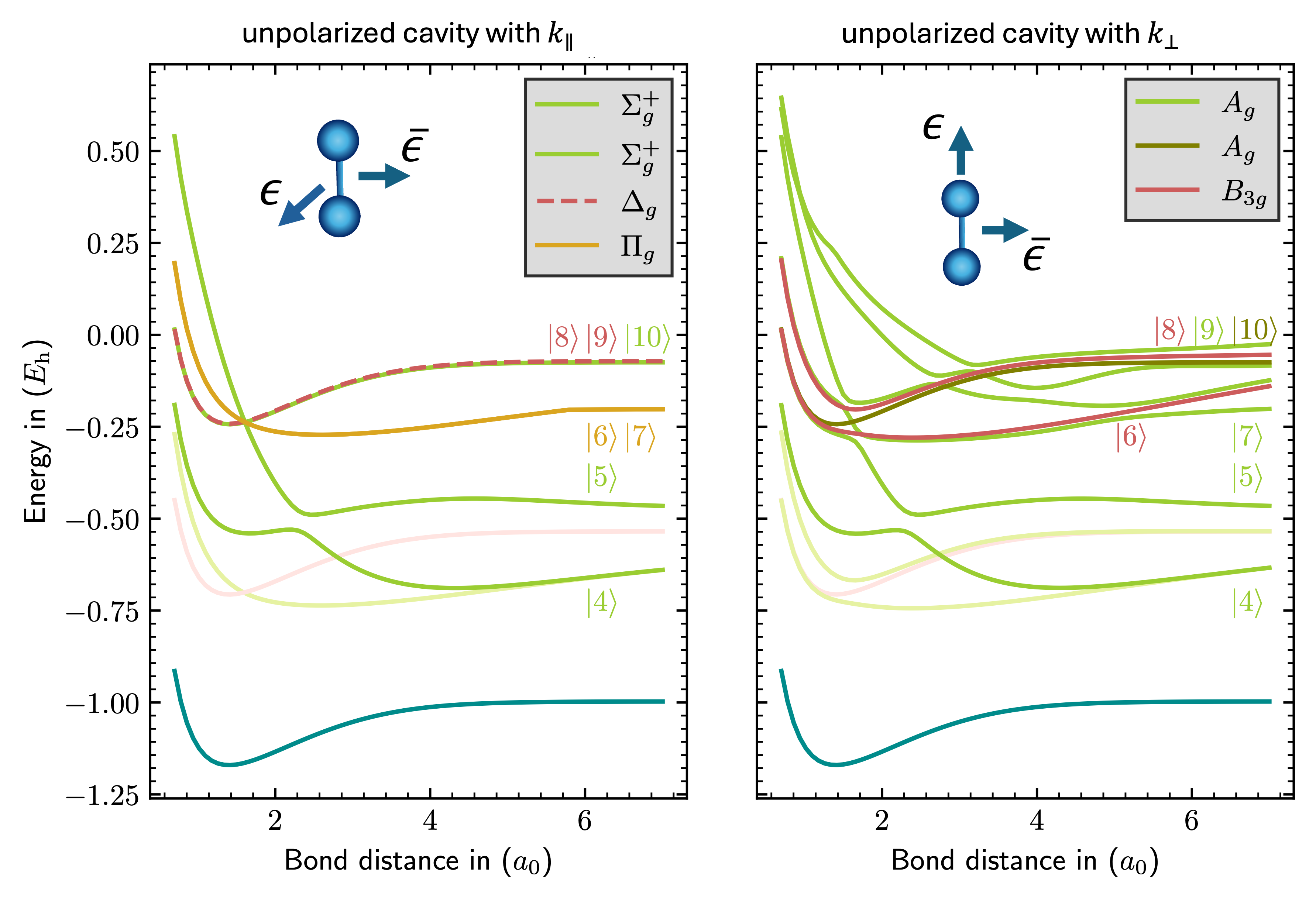}
   \end{minipage}\\
   \begin{minipage}{0.39\textwidth}
    \begin{tabular}{c c c c}
       \hline
       \hline
       state & \multicolumn{1}{|c}{sym} & \multicolumn{2}{|c}{coupling}\\
       \hline
       \multicolumn{4}{c}{parallel}   \\
       \hline
       $\ket{4}$ & $\Sigma_g^+ $ & $ \ket{\Sigma_g^+,0}  $&$ \ket{\Sigma_g^+,\Sigma_g^+}$ \\ 
       \rule{0pt}{3ex}    
       $\ket{5}$ & $\Sigma_g^+ $ & $ \ket{\Sigma_g^+,0}  $&$ \ket{\Sigma_g^+,\Sigma_g^+}$ \\ 
       \rule{0pt}{3ex}    
       \multirow{2}{*}{$\ket{6} \ket{7}$}  & \multirow{2}{*}{$\Pi_g $}  & $ \ket{\Sigma_u^+, 1} $&$ \ket{\Sigma_u^+,\Pi_u^-}$ \\
       &  & $ \ket{\Sigma_u^+, \bar{1}} $&$ \ket{\Sigma_u^+,\Pi_u^+}$ \\
       \rule{0pt}{3ex}    
       \multirow{2}{*}{$\ket{8} \ket{9}$}  & \multirow{2}{*}{$\Delta_g $}  & $ \ket{\Sigma_g^+, 2} - \ket{\Sigma_g^+, \bar{2}} $&$ \ket{\Sigma_g^+,\Delta_u^-}$ \\
       &  & $ \ket{\Sigma_g^+, 1\bar{1}} $&$ \ket{\Sigma_g^+,\Delta_u^+}$ \\
       \rule{0pt}{3ex}    
       $\ket{10}$ & $\Sigma_g^+ $ & $ \ket{\Sigma_g^+,2}+\ket{\Sigma_g^+,\bar{2}}  $&$ \ket{\Sigma_g^+,\Sigma_g^+}$ \\ 
        \hline

       \multicolumn{4}{c}{perpendicular} \\
       \hline
       $\ket{4}$ & $A_g $ & $ \ket{A_g,0}  $&$ \ket{A_g,A_g}$ \\ 
       $\ket{5}$ & $A_g $ & $ \ket{A_g,0}  $&$ \ket{A_g,A_g}$ \\ 
       $\ket{6}$ &  $B_{2g} $ &  $\ket{B_{1u},\bar{1}}  $&$ \ket{B_{1u}, B_{3u}}$ \\ 
       $\ket{7}$ &  $A_{g} $ &  $\ket{B_{1u},1}  $&$ \ket{B_{1u}, B_{1u}}$ \\ 
       $\ket{8}$ &  $B_{2g} $ &  $\ket{A_{g},1\bar{1}}  $&$ \ket{A_{g},B_{2g}}$ \\ 
       $\ket{9}$ &  $A_g $ &  $\ket{A_{g},2}  $&$ \ket{A_{g},A_g}$ \\ 
       $\ket{10}$ &  $A_g $ &  $\ket{A_{g},\bar{2}}  $&$ \ket{A_{g},A_g}$ \\ 
        \hline
        \hline
    \end{tabular}
    \end{minipage}
   \caption{Upper part: Low-lying singlet states of H$_2$ in an unpolarized cavity with the polarization vector aligned parallel (left) and perpendicular (right) with respect to the molecular axis.
   Only selected states of \textit{gerade} parity are shown.  
   The cavity frequency was set to $12.68\; eV$ and the coupling strength to $0.05 \text{a.u.}$
   Lower part: Different notations for the polaritonic states of $^1$H$_2$ in parallel as well as perpendicular orientation.
            }
   \label{fig:h2_ger_sym}
\end{figure*}

When the H$_2$ molecule is placed in an unpolarized cavity (Fig. \ref{fig:h2_un_sym}), a third excited state appears which corresponds to the additional mode, polarized perpendicular to the first. 
Practically, the unpolarized cavity appears as a combination of the two linearly polarized cavities shown in Fig. \ref{fig:h2_un_lin}. 
If the $\boldsymbol{k}$-vector is aligned parallel to the molecular axis, both polarization vectors are aligned  perpendicular to the molecule and no splitting into upper and lower polariton is obtained. 
Therefore, three excited states are obtained: one of them is mainly characterized by an electronic excitation and is of $\Sigma_u^+$ symmetry; the other two are degenerate states which are of $\Pi_u$ symmetry and have a strong photonic character. 
If the $\boldsymbol{k}$-vector is aligned perpendicularly to the molecule, one polarization vector can be aligned parallel to the molecule and therefore a coupling resulting in an upper and lower polariton is obtained. 
The other polarization vector is aligned perpendicular to the molecule and is of $B_{1u}$ symmetry which is not strongly coupled to the matter states. 
(For easier comparison to the linearly polarized cavity, the $\boldsymbol{k}$-vector is aligned parallel to the $y$-axis so that the polarization vector $\boldsymbol{\epsilon}$ is parallel to the $z$-axis and $\bar{\boldsymbol{\epsilon}}$ parallel to the $x$-axis).
A more complicated coupling scheme appears for the states of \textit{gerade} parity, shown in Figs. \ref{fig:h2_ger_lin} and \ref{fig:h2_ger_sym}.
For the linearly polarized cavity in the perpendicular orientation (Fig. \ref{fig:h2_ger_lin} left), no coupling into upper and lower polariton is observed as the two lowest-lying excited electronic states $\ket{3}$ and $\ket{4}$ are of $A_g$ symmetry. 
The mixed state $\ket{5}$ is of $B_{2g}$ symmetry and the doubly excited photonic state $\ket{6}$ is of $A_g$ symmetry.
As only states that differ by at most one photon  couple within the used truncation scheme (see Section (\ref{subsec:excited_states})), the states $\ket{A_g,0}$ and $\ket{A_g,2}$ are  allowed to cross. 
This is unusual since they are characterized by the same irreducible representation. However, including higher photonic excitations in $\hat{Q}$ should lead to a coupling of both.
Note that for the parallel orientation two additional excited states must be included as more avoided-crossings with other electronic states appear due to the symmetry reduction.
The avoided crossing appearing for the parallel orientation (right) can be understood as two separate avoided crossings, one appearing between the electronic states $\ket{\Sigma_g^+,0}$ and the $\ket{\Sigma_u^+,1}$ and the second between the photonic doubly excited ground-state $\ket{\Sigma_g^+,2}$ and the $\ket{\Sigma_u^+,1}$ state, respectively.
For the perpendicular orientation, no polaritonic splitting is observed within the investigated energy range. 
Note that this three-state avoided-crossing can only be described in the CCSD-12-SD truncation scheme as two-photon creations are required to describe the $\ket{A_g,2}$ state and would not appear in the CCSD-1-SD truncation.
\\
Comparing with the \textit{gerade} -symmetric states in an unpolarized cavity shows that these can be viewed as superpositions of states arising from two distinct linear polarization directions.
Due to the large number of excited states, the energy landscape becomes rather crowded. 
In the following, we focus on the energy region where a total of six states cross.
Only one of them, the state $\ket{5}$, is mostly electronic. 
Three states are purely photonic of which two are degenerate ($\ket{8}$ and $\ket{9}$) and are of $\Delta_g$ symmetry. The last state $\ket{10}$ is of $\Sigma_g^+$ symmetry. 
The degenerate states can be understood as a linear combination of the states  $\ket{A_g,1\bar{1}}$ and the anti-symmetric combination of $\ket{A_g,2}$ and $\ket{A_g,\bar2}$. 
The symmetric combination of $\ket{A_g,2}$ and $\ket{A_g,\bar{2}}$ is not degenerate and has an  overall $\Sigma_g^+$ symmetry.
But, as $\ket{A_g,2}$ and $\ket{A_g,\bar{2}}$ differ in more than one photon from the state $\ket{A_g,0}$, no coupling occurs, even though the states are of $\Sigma_g^+$ symmetry. 
The last two states $\ket{6}$ and $\ket{7}$ (orange) are mixed excitations. 
They are  degenerate and of $\Pi_g$ symmetry and can be understood qualitatively as  state $\ket{1}$ shifted with the cavity frequency (the excitation can either be in the mode $\alpha$ or $\bar{\alpha}$.
As all six states have either different irreducible representations or differ by more than one photon, they are allowed to cross. 

For the perpendicular orientation of the cavity, the symmetry is reduced and one of the degenerate states $\Delta_g$ becomes totally symmetric ($A_g$) while the other acquires $B_{3g}$ symmetry.  
One of the states that corresponds to $\Pi_g$ in the parallel orientation, becomes totally symmetric ($A_g$). The other is of $B_{3g}$ symmetry. 
In total, this results in an energy landscape where four states are of $A_g$  and two are of $B_{3g}$ symmetry, respectively.
The states with $B_{3g}$ symmetry are the photonic doubly excited state $\ket{A_g,2}$ and the mixed excitation $\ket{\Sigma_u^+,1}$, which form an upper and lower polariton.
From the states of $\Sigma_g^+$ symmetry, only three states are coupled, namely the states $\ket{A_g,0}$, $\ket{B_{2u},1}$ and $\ket{A_g,2}$, where the states $\ket{A_g,0}$ and $\ket{A_g,2}$ are only indirectly coupled as they differ by more than one photon.
The state $\ket{A_g,\bar{2}}$ differs by more than one photon from the other states of $A_g$ symmetry and is hence not be coupled to them within out truncation.
This results in three coupled states of $A_g$ symmetry, two coupled states of $B_{3u}$ symmetry and an uncoupled state of $A_g$ symmetry.
Again, including higher photonic excitations would lead to couplings among all $A_g$ states.

\section{Conclusions}
In this paper, we presented a generalization of polaritonic coupled-cluster (CC) theory to account for a rotationally symmetric, unpolarized cavity within the dipole approximation, such as an unpolarized Fabry-Pérot cavity. 
To maintain the $D_{\infty h}$ symmetry of the bare cavity, at least two perpendicular polarized modes must be included in the Hamiltonian. 
Calculations on the aromatic species  benzene, fluorobenzene, and azulene showed that the changes in the QED coupled cluster one-electron density vary significantly between a linearly and an unpolarized cavity. 
On the fluorobenzene molecule, we have also shown that placing a molecule in a cavity can remove electron density from the fluor-carbon bond, which might lead to the destabilization of the chemical bond.
This mechanism could potentially influence the reactivity of the molecule.
Furthermore, for the one-electron density of the azulene molecule, QED-CC calculations were  compared to QEDFT results. 
These have revealed some agreement in the qualitative shift of the one-electron density, but also show clear differences which should be investigated further.
\\
An efficient exploitation of  point-group symmetry based on the direct product decomposition has been presented which leads to a significant speed up of the QED-CC and QED-EOM-CC calculations. 
It also allows the characterization of excited states according to the irreducible representations and allows to predict between which states a Rabi splitting is formed. 
\\
Studies on the H$_2$ molecule showed rich coupling patterns, where the exploitation of point-group symmetry allowed the characterization of polaritonic excited states and enabled their targeted calculation. 
It has also been shown for the H$_2$ molecule 
 that the unpolarized cavity can
be qualitatively understood as 
as effectively representing a combination of two orthogonal linear polarizations.
\\
Future work should address the effect of mass renormalization and how the matter system changes when including more cavity modes.

\begin{acknowledgement}
The authors thank Michael Ruggenthaler for helpful discussions.
This work was funded by the Deutsche Forschungsgemein-
schaft (DFG, German Research Foundation)Project-ID
429529648TRR 306 QuCoLiMa (“Quantum Cooperativity
of Light and Matter”).  

\end{acknowledgement}

\appendix
\section{Transformation to Complex Polarization Vectors}
\label{sec:circ}
It should be  illustrated that the cavity is neither linearly or circularly polarized by transforming the displacement field $\hat{\boldsymbol{D}}$ in a basis of complex polarization vectors. 
Using the unitary operator
\begin{equation}
  \hat{U}_\varphi  = \text{exp} \Big[ i \varphi \sum_\alpha^{N_\text{cav}/2} ( \hat{\alpha}^\dagger \hat{\bar{\alpha}} + \hat{\bar{\alpha}}^\dagger \hat{\alpha} ) \Big]   \; .
   \label{}
\end{equation}
the elementary operators can be transformed as 
\begin{equation}
   \begin{split}
        \hat{\alpha}  & \rightarrow \hat{\alpha}  \cos\varphi + i \hat{\bar{\alpha}} \sin\varphi\\
        \hat{\bar{\alpha}}   & \rightarrow \hat{\bar{\alpha}}  \cos\varphi + i \hat{\alpha} \sin\varphi\\
        \hat{\alpha}^\dagger   & \rightarrow \hat{\alpha}^\dagger  \cos\varphi - i \hat{\bar{\alpha}}^\dagger \sin\varphi\\
        \hat{\bar{\alpha}}^\dagger   & \rightarrow \hat{\bar{\alpha}}^\dagger  \cos\varphi - i \hat{\alpha}^\dagger \sin\varphi\;.\\
   \end{split}
   \label{eq:circ_modes}
\end{equation}
This corresponds to a transformation into a basis of elliptical polarization vectors and can be made explicit when inserting the elementary operators (\ref{eq:circ_modes}) in the displacement field (\ref{eq:efield_displace}).
E.g. using the angle $\varphi=\pi/2$, the polarization vectors transform as
\begin{equation}
   \begin{split}
     \boldsymbol{\epsilon}^\prime & =  \frac{1}{\sqrt{2}}(\boldsymbol{\epsilon} + i \bar{\boldsymbol{\epsilon}})\;,\\
     \bar{\boldsymbol{\epsilon}}^\prime & =  \frac{1}{\sqrt{2}}(\bar{\boldsymbol{\epsilon}} + i \boldsymbol{\epsilon} )\;,
   \end{split}
   \label{}
\end{equation}
hence, they become circularly polarized. 
This should illustrate that the polarization vectors only serve as a basis for the representation of the EM field.
Whether real or complex valued polarization vectors polarization vectors are used does not change the underlying physical properties.

\begin{suppinfo}
\newpage
\onecolumn

\begin{figure}[h!]
   \centering
   \includegraphics[width=0.5\linewidth,trim=4 4 4 4,clip]{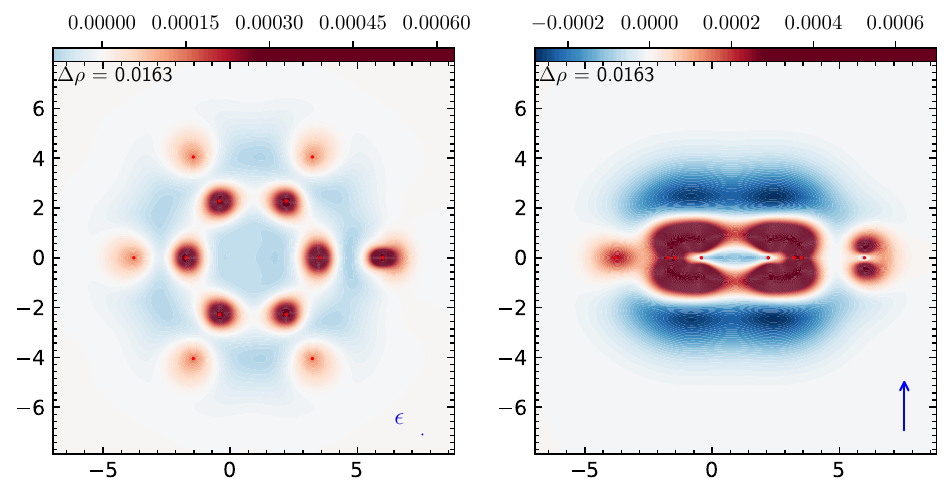}
   \includegraphics[width=0.5\linewidth,trim=4 4 4 4,clip]{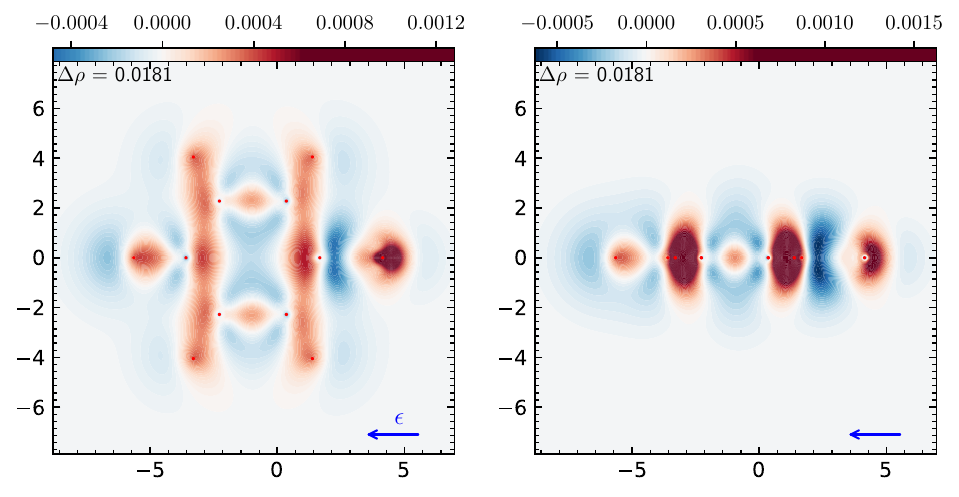}
   \caption{Correlated one-electron density differences for fluorobenzene in two orientations within a linearly polarized cavity. 
            }
   \label{fig:fluor_benz_difference_lin}
\end{figure}
\newpage

\begin{longtable}{c c c}
\caption{Geometry for the benzene molecule in $a_0$.}
\label{tab:h2_un_lin}\\
\hline
\hline
Atom center &  \multicolumn{2}{c}{Cartesian geometry in $a_0$} \\
\hline
H &   4.04983121690698   &    2.33848216048779  \\
C &   2.27711795932538   &    1.31460686449430  \\
C &   2.27711795932538   &   -1.31460686449430  \\
H &   4.04983121690698   &   -2.33848216048779  \\
C &   0.00000000000000   &   -2.62905343248059  \\
H &   0.00000000000000   &   -4.67606322887373  \\
C &  -2.27711795932538   &   -1.31460686449430  \\
H &  -4.04983121690698   &   -2.33848216048779  \\
C &  -2.27711795932538   &    1.31460686449430  \\
H &  -4.04983121690698   &    2.33848216048779  \\
C &   0.00000000000000   &    2.62905343248059  \\
H &   0.00000000000000   &    4.67606322887373  \\
\hline
\hline
\end{longtable}

\begin{longtable}{c c c}
\caption{Geometry for the azulene molecule in $a_0$.}
\label{tab:h2_un_lin}\\
\hline
\hline
Atom center &  \multicolumn{2}{c}{Cartesian geometry in $a_0$} \\
\hline
C  & -1.26866806853591 &  1.92438989279682 \\  
C  & -1.59699010340496 &  0.56261557814130 \\
H  & -2.66893282091365 &  0.33204619125468 \\
C  & -0.75099634964616 & -0.54694244712311 \\
C  & -1.15211338363022 & -1.89765202634847 \\
C  &  0.75099634964616 & -0.54694244712311 \\
C  &  1.15211338363022 & -1.89765202634847 \\
C  &  1.59699010340496 &  0.56261557814130 \\
H  &  2.66893282091365 &  0.33204619125468 \\
C  &  1.26866806853591 &  1.92438989279682 \\
H  &  2.11573594524611 &  2.61742512270539 \\
C  &  0.00000000000000 & -2.70561721020542 \\
H  & -2.11573594524611 &  2.61742512270539 \\
H  & -2.18544728279424 & -2.24575344665612 \\
H  &  2.18544728279424 & -2.24575344665612 \\
H  &  0.00000000000000 & -3.79752591434090 \\
C  &  0.00000000000000 &  2.51809656766395 \\
H  &  0.00000000000000 &  3.61381384682047 \\
\hline
\hline
\end{longtable}

\begin{longtable}{c c c}
\caption{Geometry for the fluorobenzene molecule in $a_0$.}
\label{tab:h2_un_lin}\\
\hline
\hline
Atom center &  \multicolumn{2}{c}{Cartesian geometry in $a_0$} \\
\hline
F &  0.00000000000000  & -5.1000000000000  \\ 
H &  4.04983121690698  &  2.33848216048779 \\ 
C &  2.27711795932538  &  1.31460686449430 \\ 
C &  2.27711795932538  & -1.31460686449430 \\ 
H &  4.04983121690698  & -2.33848216048779 \\ 
C &  0.00000000000000  & -2.62905343248059 \\ 
C & -2.27711795932538  & -1.31460686449430 \\ 
H & -4.04983121690698  & -2.33848216048779 \\ 
C & -2.27711795932538  &  1.31460686449430 \\ 
H & -4.04983121690698  &  2.33848216048779 \\ 
C &  0.00000000000000  &  2.62905343248059 \\ 
H &  0.00000000000000  &  4.67606322887373 \\ 
\hline
\hline
\end{longtable}

\begin{table}
\caption{Benzene molecule in a linear and unpolarized cavity with $\lambda=0.05\text{ a.u}$. A cc-PVTZ basis was used. Table contains all energies in units of eV.}
\label{tab:h2_un_lin}
\begin{tabular}{c c c c c c c}
\hline
\hline
$\omega_\text{cav} \;\text{(eV)}$ & $\epsilon_x$ & $\epsilon_y$& $\epsilon_z$& $k_x$& $k_y$ & $k_z$\\
\hline
0.00027211 &  0.73493522 & 0.73492376 & 0.52666601 & 1.2614862  & 1.26149764  & 1.47040564 \\
1.36082831 &  0.68098129 & 0.68097185 & 0.50066712 & 1.18160194 & 1.18161142  & 1.3626517  \\
2.72138452 &  0.63559943 & 0.63559172 & 0.47742223 & 1.11303745 & 1.11304522  & 1.27204594 \\
4.08194072 &  0.59669061 & 0.59668437 & 0.45647777 & 1.05323838 & 1.05324469  & 1.19436365 \\
5.44249692 &  0.56283778 & 0.5628328  & 0.43748297 & 1.00043719 & 1.00044225  & 1.12676687 \\
6.80305312 &  0.53303693 & 0.533033   & 0.42015996 & 0.95335277 & 0.95335677  & 1.06725007 \\
8.16360932 &  0.50654976 & 0.50654672 & 0.40428443 & 0.9110234  & 0.91102651  & 1.01434056 \\
\hline
\hline
\end{tabular}
\end{table}

\newpage
\begin{longtable}{c c c c c c c}
\caption{H$_2$ molecule in an unpolarized cavity with $\lambda=0.05\text{ a.u.}$ and $\omega=0.466\text{ E}_h$. The $k$-vector is oriented parallel to the molecular axis. A cc-pVTZ basis was used. Table contains all energies in units of eV.}
\label{tab:h2_un_lin}\\
\hline
\hline
$R \;\text{in}\; (a_0)$ & $\epsilon_\perp(A_g)$ & $\epsilon_\perp(B_{2u})$  & $\epsilon_\perp(B_{3u})$ & $\epsilon_\parallel(A_g)$ & $\epsilon_\parallel(B_{1u})$& $\epsilon_\parallel(B_{1u})$   \\
\hline
0.7        & -0.91518109 &-0.45223356 & -0.45044443 & -0.91510052 &  -0.27106995 & -0.26672831  \\  
0.77974684 & -0.99773758 &-0.53533116 & -0.53308245 & -0.99764795 &  -0.37325843 & -0.36830318  \\ 
0.85949367 & -1.05622458 &-0.59449035 & -0.59165229 & -1.05612457 &  -0.45049873 & -0.44479586  \\ 
0.93924051 & -1.09757565 &-0.63668307 & -0.63308717 & -1.09746419 &  -0.50980075 & -0.50318091  \\ 
1.01898734 & -1.12647593 &-0.66664802 & -0.66207199 & -1.12635222 &  -0.55596969 & -0.54821096  \\ 
1.09873418 & -1.14618722 &-0.68772238 & -0.68186827 & -1.14605065 &  -0.59234153 & -0.58314444  \\
\hline
\hline
\end{longtable}

\newpage
\begin{longtable}{c c c c c}
  \caption{H$_2$ molecule in an unpolarized cavity with $\lambda=0.05\text{ a.u.}$ and $\omega=0.466\text{ E}_h$. The $k$-vector is oriented parallel to the molecular axis. Ground state and excited state with \textit{ungerade} symmetry. A cc-pVTZ basis was used. All energies in eV.}
\label{tab:h2_un_lin}\\
\hline
\hline
$R \;\text{in}\; (a_0)$ & $k_\parallel(A_g)$ & $k_\parallel(B_{1u})$  & $k_\parallel(B_{2u})$  & $k_\parallel(B_{3u})$   \\
\hline
0.7        & -0.91435815 &   -0.27014046 & -0.44962085 & -0.44962085  \\  
0.77974684 & -0.99689111 &   -0.37232121 & -0.53223536 & -0.53223536  \\ 
0.85949367 & -1.0553554  &   -0.44955145 & -0.5907825  & -0.5907825   \\ 
0.93924051 & -1.09668459 &   -0.50884111 & -0.63219553 & -0.63219553  \\ 
1.01898734 & -1.12556386 &   -0.55499687 & -0.66115935 & -0.66115935  \\ 
1.09873418 & -1.14525498 &   -0.59135616 & -0.68093548 & -0.68093548  \\
\hline
\hline
\end{longtable}

\newpage
\begin{longtable}{c c c c c}
\caption{H$_2$ molecule in an unpolarized cavity with $\lambda=0.05\text{ a.u.}$ and $\omega=0.466\text{E}_h$. The $k$-vector is oriented perpendicular to the molecular axis. A cc-pVTZ basis was used. All energies in eV.}
\label{tab:h2_un_lin}\\
\hline
\hline
$R \;\text{in}\; (a_0)$ & $k_\perp(A_g)$ & $k_\perp(B_{1u})$  & $k_\perp(B_{2u})$  & $k_\perp(B_{3u})$    \\
\hline
0.7         & -0.91427717 & -0.44953953 &  -0.45140329 & -0.26580853 \\  
0.77974684  & -0.99680105 & -0.5321449  &  -0.53447611 & -0.3673769  \\ 
0.85949367  & -1.05525494 & -0.59068157 &  -0.5936111  & -0.44386094 \\ 
0.93924051  & -1.09657267 & -0.63208306 &  -0.6357802  & -0.5022353  \\ 
1.01898734  & -1.12543966 & -0.66103452 &  -0.66572203 & -0.54725421 \\ 
1.09873418  & -1.14511791 & -0.6807977  &  -0.68677374 & -0.58217764 \\
\hline
\hline
\end{longtable}

\newpage
\begin{longtable}{c c c c c}
\caption{H$_2$ molecule in an linearly cavity with $\lambda=0.05\text{ a.u.}$ and $\omega=0.466\text{E}_h$. The $\epsilon$-vector is oriented perpendicular to the molecular axis. A cc-pVTZ basis was used. All energies in eV.}
\label{tab:h2_ger_lin_ex}\\
\hline
\hline
$R \;\text{in}\; (a_0)$ & $\epsilon_\perp(A_g)$ & $\epsilon_\perp(A_g)$  & $\epsilon_\perp(A_g)$  & $\epsilon_\perp(B_{1g})$    \\
\hline
0.7        & -0.19127305 &  0.53931945 &  0.01445393 &   0.19380488 \\
0.77974684 & -0.28645546 &  0.42735493 & -0.06824342 &   0.0915989  \\
0.85949367 & -0.35669473 &  0.3353046  & -0.12687232 &   0.01433929 \\
0.93924051 & -0.4089465  &  0.25399791 & -0.16836578 &  -0.04498463 \\
1.01898734 & -0.44793252 &  0.17831019 & -0.19740851 &  -0.09117729 \\
1.09873418 & -0.47695862 &  0.10701739 & -0.2172619  &  -0.12757403 \\
\hline
\hline
\end{longtable}

\newpage
\begin{longtable}{c c c c c c c}
\caption{H$_2$ molecule in an linearly cavity with $\lambda=0.05\text{ a.u.}$ and $\omega=0.466\text{ E}_h$. The $\epsilon$-vector is oriented parallel to the molecular axis. A cc-pVTZ basis was used. All energies in eV.}
\label{tab:h2_ger_lin_ez}\\
\hline
\hline
$R \;\text{in}\; (a_0)$ & $\epsilon_\parallel(A_g)$ & $\epsilon_\parallel(A_g)$  & $\epsilon_\parallel(A_g)$  & $\epsilon_\parallel(A_{g})$& $\epsilon_\parallel(A_{g})$& $\epsilon_\parallel(A_{g})$    \\
\hline
 0.7        &   -0.19183627 &   0.01225997 &   0.205058   &   0.53717112 &   0.64388013 &   0.73859121 \\
 0.77974684 &   -0.28703266 &  -0.07095505 &   0.10321834 &   0.42619922 &   0.54087441 &   0.59608599 \\
 0.85949367 &   -0.35727893 &  -0.1302351  &   0.02640175 &   0.33498165 &   0.46034301 &   0.48966146 \\
 0.93924051 &   -0.40953425 &  -0.17254888 &  -0.03236953 &   0.25424175 &   0.38390015 &   0.42381552 \\
 1.01898734 &   -0.44852545 &  -0.2026304  &  -0.0778575  &   0.17902297 &   0.31907492 &   0.3783662  \\
 1.09873418 &   -0.47756111 &  -0.2238106  &  -0.11334958 &   0.10826645 &   0.26836026 &   0.34001318 \\
\hline
\hline
\end{longtable}

\newpage
\begin{longtable}{c c c c c c c}
\caption{H$_2$ molecule in an unpolarized cavity with $\lambda=0.05\text{ a.u.}$ and $\omega=0.466\text{ E}_h$. The $k$-vector is oriented parallel to the molecular axis. A cc-pVTZ basis was used. All energies in eV.}
\label{tab:h2_ger_unp_kz}\\
\hline
\hline
$R \;\text{in}\; (a_0)$ & $k_\parallel(A_g)$ & $k_\parallel(A_g)$  & $k_\parallel(A_g)$  & $k_\parallel(B_{1g})$ & $k_\parallel(B_{2g})$& $k_\parallel(B_{3g})$    \\
\hline
 0.7        &  -0.19052919 &   0.53991143 &   0.01438372 &  0.0152772  &   0.19499212 &   0.19499212 \\
 0.77974684 &  -0.28570478 &   0.42803183 &  -0.06835386 & -0.06739665 &   0.09279145 &   0.09279145 \\
 0.85949367 &  -0.35593667 &   0.33609188 &  -0.12702539 & -0.12600287 &   0.01554055 &   0.01554055 \\
 0.93924051 &  -0.40818074 &   0.25491034 &  -0.16856371 & -0.16747451 &  -0.04377123 &  -0.04377123 \\
 1.01898734 &  -0.44715887 &   0.17932977 &  -0.19765333 & -0.19649629 &  -0.08995044 &  -0.08995044 \\
 1.09873418 &  -0.47617697 &   0.10810621 &  -0.21755542 & -0.21632959 &  -0.12633441 &  -0.12633441 \\
\hline
\hline
\end{longtable}

\newpage
\begin{landscape}
\begin{longtable}{c c c c c c c c}
\caption{H$_2$ molecule in an unpolarized cavity with $\lambda=0.05\text{ a.u.}$ and $\omega=0.466\text{ E}_h$. The $k$-vector is oriented perpendicular to the molecular axis. A cc-pVTZ basis was used. All energies in eV.}
\label{tab:h2_ger_unp_kx}\\
\hline
\hline
$R \;\text{in}\; (a_0)$ & $k_\perp(A_g)$ & $k_\perp(A_g)$  & $k_\perp(A_g)$  & $k_\perp(A_g)$ & $k_\perp(A_g)$& $k_\perp(A_g)$ & $k_\perp(A_g)$\\
\hline
0.700000000 & -0.191103789e &  0.010667245 &  0.208345536     &  0.537775138     &  0.014468647 &  0.614620187 &  0.644941196 \\
0.779746835 & -0.286294169e & -0.072966900 &  0.106945910     &  0.426880508     & -0.068259508 &  0.514077731 &  0.541956570 \\
0.859493671 & -0.356533972e & -0.132764474 &  0.030675214     &  0.335769201     & -0.126920244 &  0.438360000 &  0.461423392 \\
0.939240506 & -0.408782563e & -0.175713454 & -0.027426242     &  0.255152581     & -0.168446650 &  0.380155916 &  0.385345735 \\
1.01898734e & -0.447766915e & -0.206568787 & -0.721040283     &  0.180039723     & -0.197523508 &  0.319736690 &  0.336366133 \\
1.09873418e & -0.476795745e & -0.228681190 & -0.106630018     &  0.109352969     & -0.217412203 &  0.269057831 &  0.301772866 \\
\hline
\hline
\end{longtable}

\end{landscape}

\newpage
\begin{longtable}{c c c}
\caption{H$_2$ molecule in an unpolarized cavity with $\lambda=0.05\text{ a.u.}$ and $\omega=0.466\text{ E}_h$. The $k$-vector is oriented perpendicular to the molecular axis. A cc-pVTZ basis was used. All energies in eV.}
\label{tab:h2_ger_unp_kx_2}\\
\hline
\hline
$R \;\text{in}\; (a_0)$ & $k_\perp(B_{3g})$ & $k_\perp(B_{3g})$    \\
\hline
 0.7        &  0.20217202 &   0.01246896 \\
 0.77974684 &  0.10047302 &  -0.07074514 \\
 0.85949367 &  0.02384742 &  -0.13002296 \\
 0.93924051 & -0.03467919 &  -0.17233398 \\
 1.01898734 & -0.07985986 &  -0.20241301 \\
 1.09873418 & -0.11495652 &  -0.2235911  \\
\hline
\hline
\end{longtable}
\twocolumn

\end{suppinfo}

\bibliography{achemso-demo}

\end{document}